\documentclass[sigconf]{acmart}
\usepackage[utf8]{inputenc}
\usepackage[ruled]{algorithm2e} 
\usepackage{enumitem}
\usepackage{amsmath}
\usepackage{graphicx}
\usepackage{capt-of}

\usepackage{color}
\usepackage{colortbl}
\usepackage{soul}          
\usepackage[draft,inline,nomargin]{fixme}
\usepackage{multirow}
\usepackage{epstopdf}
\usepackage{booktabs}
\usepackage{array}
\usepackage{url}
\input{glyphtounicode}
\usepackage[T1]{fontenc}
\usepackage{subcaption}
\usepackage{tabularx}
\usepackage{ltxtable}
\usepackage{blindtext}
\usepackage{graphicx}
\usepackage{wrapfig,lipsum,booktabs}
\usepackage{booktabs}
\usepackage{array}
\usepackage{svg}
\usepackage{arydshln}
\usepackage{booktabs}
\usepackage{stackengine}
\def\mybar[#1]#2{
  {\color{black}\rule[0.1ex]{#1mm}{5pt}} #2}
\usepackage{acmart-taps}

\SetAlFnt{\small}
\SetAlCapFnt{\small}
\SetAlCapNameFnt{\small}
\SetAlCapHSkip{0pt}
\IncMargin{-\parindent}

\newcommand\para[1]{{\vspace{5pt} \noindent {\bf #1}}}

\newcolumntype{L}[1]{>{\raggedright\let\newline\\\arraybackslash\hspace{0pt}}m{#1}}
\newcolumntype{C}[1]{>{\centering\let\newline\\\arraybackslash\hspace{0pt}}m{#1}}
\newcolumntype{R}[1]{>{\raggedleft\let\newline\\\arraybackslash\hspace{0pt}}m{#1}}

\def\mybar[#1]#2{
  \hspace{5mm}{\color{black}\rule[0.1ex]{#1mm}{4pt}} #2}

\copyrightyear{2024}
\acmYear{2024}
\setcopyright{rightsretained}
\acmConference[CHI '24]{Proceedings of the CHI Conference on Human Factors in Computing Systems}{May 11--16, 2024}{Honolulu, HI, USA}
\acmBooktitle{Proceedings of the CHI Conference on Human Factors in Computing Systems (CHI '24), May 11--16, 2024, Honolulu, HI, USA}
\acmDOI{10.1145/3613904.3642680}
\acmISBN{979-8-4007-0330-0/24/05}

\begin{document}

\title{MoodCapture: Depression Detection Using In-the-Wild Smartphone Images}

\author{Subigya Nepal}
\orcid{0000-0002-4314-9505}
\authornote{\textbf{Both authors contributed equally to this research.}}
\author{Arvind Pillai}
\orcid{0000-0002-2489-1130}
\authornotemark[1]
\affiliation{%
  \institution{Dartmouth College}
  \streetaddress{15 Thayer Dr}
  \city{Hanover}
  \state{New Hampshire}
  \country{USA}
  \postcode{03755}
}

\author{Weichen Wang}
\orcid{0000-0001-6738-9944}
\affiliation{%
  \institution{Dartmouth College}
  \streetaddress{15 Thayer Dr}
  \city{Hanover}
  \state{New Hampshire}
  \country{USA}
  \postcode{03755}
}
\author{Tess Griffin}
\orcid{0000-0001-5462-575X}
\affiliation{%
  \institution{Dartmouth College}
  \city{Hanover}
  \state{New Hampshire}
  \country{USA}
}
\author{Amanda C. Collins}
\orcid{0000-0002-8258-2272}
\affiliation{%
  \institution{Dartmouth College}
  \city{Hanover}
  \state{New Hampshire}
  \country{USA}
}
\author{Michael Heinz}
\orcid{0000-0003-0866-0508}
\affiliation{%
  \institution{Dartmouth College}
  \city{Hanover}
  \state{New Hampshire}
  \country{USA}
}
\author{Damien Lekkas}
\orcid{0000-0002-6995-9223}
\affiliation{%
  \institution{Dartmouth College}
  \city{Hanover}
  \state{New Hampshire}
  \country{USA}
}
\author{Shayan Mirjafari}
\orcid{0000-0002-7165-2781}
\affiliation{%
  \institution{Dartmouth College}
  \city{Hanover}
  \state{New Hampshire}
  \country{USA}
}
\author{Matthew Nemesure}
\orcid{0000-0002-2369-600X}
\affiliation{%
  \institution{Dartmouth College}
  \city{Hanover}
  \state{New Hampshire}
  \country{USA}
}
\author{George Price}
\orcid{0000-0002-9164-4973}
\affiliation{%
  \institution{Dartmouth College}
  \city{Hanover}
  \state{New Hampshire}
  \country{USA}
}
\author{Nicholas C. Jacobson}
\orcid{0000-0002-8832-4741}
\affiliation{%
  \institution{Dartmouth College}
  \city{Hanover}
  \state{New Hampshire}
  \country{USA}
}
\author{Andrew T. Campbell}
\orcid{0000-0001-7394-7682}
\affiliation{%
  \institution{Dartmouth College}
  \city{Hanover}
  \state{New Hampshire}
  \country{USA}
}

\begin{abstract}
MoodCapture presents a novel approach that assesses depression based on images automatically captured from the front-facing camera of smartphones as people go about their daily lives. We collect over 125,000 photos in the wild from N=177 participants diagnosed with major depressive disorder for 90 days. Images are captured naturalistically while participants respond to the PHQ-8 depression survey question: \textit{``I have felt down, depressed, or hopeless''}. Our analysis explores important image attributes, such as angle, dominant colors, location, objects, and lighting. We show that a random forest trained with face landmarks can classify samples as depressed or non-depressed and predict raw PHQ-8 scores effectively. Our post-hoc analysis provides several insights through an ablation study, feature importance analysis, and bias assessment. Importantly, we evaluate user concerns about using MoodCapture to detect depression based on sharing photos, providing critical insights into privacy concerns that inform the future design of in-the-wild image-based mental health assessment tools.
\end{abstract}



\maketitle
\renewcommand{\shortauthors}{Nepal and Pillai et al.}
\section{Introduction}
\label{sec:intro} 

Today, most people automatically unlock their phones using camera biometrics and face recognition. The front-facing camera quietly captures glimpses of users' faces tens to hundreds of times daily, week in and week out. Unlike selfies, these in-the-moment images capture authentic, unguarded facial expressions, free from biases such as social desirability and self-presentation. We envision a future where AI processes these unguarded facial images on the phone in real-time using deep learning, assessing the user's mood without needing the images to leave the device, thus safeguarding privacy. This low-burden, continuous approach to depression assessment and detection will significantly alter how mental health is passively assessed, enabling early detection of depression, timely intervention, and constant evaluation of individuals at risk. This paper discusses the first steps toward realizing this vision.

Depression is a complex and pervasive mental health issue affecting millions of people worldwide. According to the World Health Organization (WHO), over 264 million people suffer from depression~\cite{WHO2023}, making it a leading cause of disability and a major contributor to the overall global burden of disease. The consequences of depression extend beyond emotional distress~\cite{Rottenberg2005}, significantly impacting physical health~\cite{Frerichs1982, Olver2013}, social relationships~\cite{Santini2015}, and occupational functioning~\cite{Deady2021}. In severe cases, depression can lead to suicide, accounting for nearly 800,000 deaths each year~\cite{Amaltinga2020, Bertolote2003}. The need for early detection and intervention in depression is critical, as timely identification of the condition allows individuals to access appropriate treatment and support, thereby improving clinical outcomes and reducing the risk of long-term complications~\cite{Saarni2007-sd, Fergusson2002-nc}. 

Smartphones offer an opportunity to explore alternative approaches for depression detection that are more objective, unobtrusive, and continuous. The vast amounts of data generated through daily smartphone usage, including images, text messages, and social media interactions, provide a rich and ecologically valid source of information that can be utilized to gain insights into an individuals' mental state. Consequently, several studies have made use of smartphone sensing data to assess depression~\cite{wang2018tracking, chikersal2021detecting}.

\begin{figure*}[ht!]
    \centering
\includegraphics[width=\textwidth]{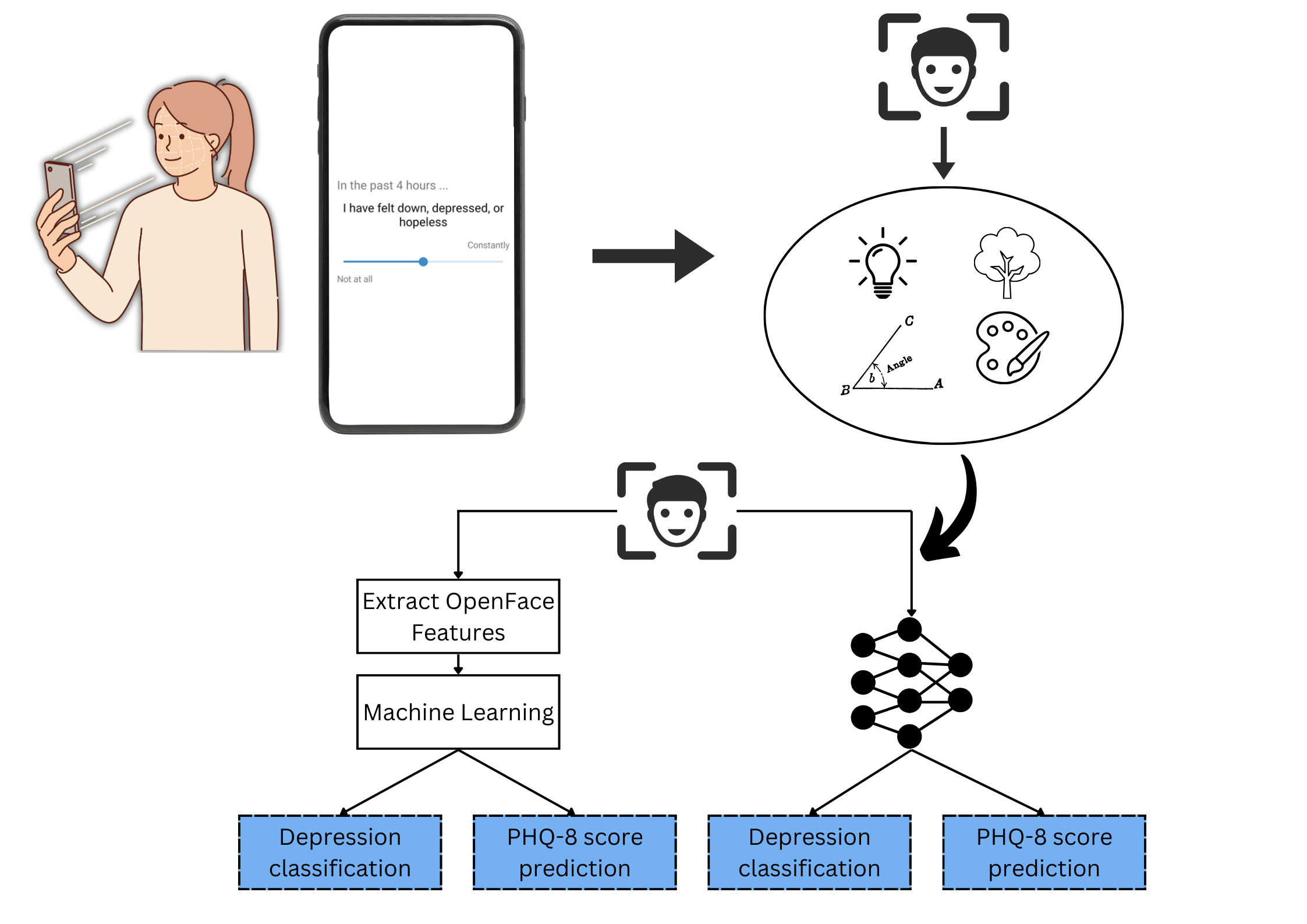}
    \caption{MoodCapture Framework: Users answer the PHQ-8 depression survey questions using the MoodCapture Android App while the app takes bursts of photos using the front-facing camera on the smartphone (top-left). Image characteristics are analysed using factors, such as, illumination, indoor vs. outdoors, phone angle, dominant image color, and background objects (top-right). Given that raw images compromise privacy, these characteristics provide insights into the types of features our machine learning and deep learning model infer. Finally, OpenFace features are extracted to train machine learning models, while raw images are used to train deep learning models (bottom). Depression classification is a binary predictor that classifies an image as depressed or not depressed, whereas PHQ-8 score prediction is a regression model that predicts raw PHQ-8 scores.}
    \label{fig:lifesense}
     \Description{Figure shows the overview diagram of the study. We take participant's pictures when they answer an item from the PHQ-8 survey, we extract several image characteristics from these images and then we train ML and DL models to predict depression using these images.}
\end{figure*}

Most of the prior research utilizing facial images to detect depression focus on capturing these images in controlled settings, where individuals may be instructed to perform specific actions \cite{Kong2022, Lee2022, Liu2022, Francese2022}. These face features are not authentic as they are performative and are influenced by biases such as social desirability and self-presentation. Furthermore, traditional methods such as clinical assessments and subjective self-reports are time-consuming and affected by recall bias. Advances in smartphone cameras offer a solution to address these disadvantages. To this end, we present MoodCapture, a novel approach to collect \textit{in-the-wild} face images and self-reported depression symptoms in natural, everyday environments using smartphones. The resulting face images capture authentic and unguarded facial expressions. Thus, minimizing the influence of self-awareness on emotions and enhancing the credibility of our data. By using such naturalistic images for analysis and training machine and deep learning models, we can better understand intricate patterns associated with depression. Ultimately, insights from our work can be used to create accurate, efficient, and personalized tools for depression detection.

Our paper contributes to the growing intersection of Human-Computer Interaction (HCI) research and mental health assessment by investigating the potential of machine learning and deep learning models trained using in-the-wild smartphone images for identifying depressive symptoms. We collected over 125,000 images from N=177 participants diagnosed with major depressive disorder over three months, utilizing 87 distinct types of Android devices owned by users in the study. On average, each participant provided six photos per day, creating a varied and extensive dataset. We comprehensively analyze various image characteristics obtained from these images captured in-the-wild. We evaluate the performance of machine learning and deep learning models trained to predict depression based on these images, as shown in Figure \ref{fig:lifesense}. At the end of the study period, we assess user acceptance by inquiring about participants' comfort levels and privacy concerns in sharing their photos for mental health assessment purposes. Therefore, our research aims to foster the development of more ethically sound mental health assessment and intervention tools. The contributions of our work are as follows:

\begin{itemize}
    \item We develop a passive-sensing image-based mobile app called MoodCapture that automatically collects \textit{in-the-wild} smartphone images from participants' front-facing cameras, ensuring an unobtrusive data collection process and maintaining user privacy. Compared to prior studies, our application captures front-facing photos in-the-moment, resulting in naturalistic images with authentic emotions. Our app provides valuable insights for future in-the-wild studies.
    \item We analyze different image characteristics such as illumination, location, phone angle, background color, and objects, providing insights into the visual properties of smartphone images. For example, majority of the images were taken indoors in well lit environments. These properties are crucial for model training and informs HCI practitioners about the environmental conditions in user interactions.
    \item We evaluate the performance of several machine learning and deep learning models for depression detection and PHQ-8 score prediction. A random forest trained with 3D face landmarks demonstrates the feasibility of analyzing depression from in-the-wild smartphone images, resulting in a balanced accuracy of 0.60, Matthew's Correlation Coefficient (MCC) of 0.14 and Mean Absolute Error (MAE) of 130.31 (a 6\% improvement over baseline on a 0-800 scale). Furthermore, we identify important features providing useful insights for HCI design. 
    \item We report on user acceptance with respect to the comfort levels of the participants in sharing their photos for mental health assessment, providing valuable insights into privacy concerns that inform the future design of in-the-wild image-based mental health assessment tools.
\end{itemize}

In addition to its relevance to the HCI community, our MoodCapture study contributes to affective computing, which deals with recognizing, interpreting, and simulating human emotions. By leveraging computational methods and machine learning models to interpret emotional cues from images, our research contributes to the understanding and development of affective computing within the HCI field. Furthermore, our study has tangible, real-world implications, such as the potential benefits of early depression detection, timely interventions, improved clinical outcomes, and overall well-being for individuals.

This paper is structured as follows: Section 2, presents related works in depression detection and work that uses smartphone images. Section 3 details the MoodCapture study, participant demographics, and the analysis we perform to identify image characteristics and to detect depression. Section 4, discusses our results, while Section 5 describes the ethical considerations and user acceptance study. Section 6 discusses the study findings and its implications. Finally, Section 7 and Section 8, discuss the limitations of the study and provide some concluding remarks, respectively.

\section{Related Work}
\label{sec:related_work}
In this section, we delve into the pertinent literature, examining the key studies and developments in the field that inform the foundation of our MoodCapture research.
\subsection{{Smartphones and Mental Health}}
Depression has been traditionally diagnosed through clinical interviews or self-reporting questionnaries such as the Beck Depression Inventory (BDI) \cite{beck1987beck} and the Hamilton Depression Rating Scale (HDRS) \cite{Kobak2010}. However, these tools are affected by the individuals' subjective recollections, social desirability bias, mental health stigmas, or the person's diminished self-awareness~\cite{Hunt2003, Schomerus2012, AriasdelaTorre2020}. Therefore, the pervasive, objective, and continuous nature of multifaceted smartphone data makes it an ideal candidate for unobtrusive depression detection. Many studies evaluate patterns in call logs, text messages, GPS coordinates, and overall smartphone activity, to gain insights into behavioral shifts, social engagement frequencies, and alterations in daily routines, all of which can serve as indicators of deteriorating mental health~\cite{wang2018tracking, xu2019leveraging, chikersal2021detecting, 10.1145/3491102.3502043}. Other modalities such as speech have also gained traction in evaluating mental health symptoms such as suicidal ideation~\cite{belouali2021acoustic, pillai2024investigating}. The growth of social media platforms provides ways to harness user-generated content for depression detection. In particular, analytical approaches using text and images have been applied to content from platforms like Facebook and Instagram. For instance, the linguistic attributes of posts can shed light on a user's emotional state, sentiment, and overall mental well-being~\cite{de2013predicting, chancellor2020methods}. Moreover, machine learning algorithms have been employed to decipher patterns and indicators of depression from visual content shared on these platforms. Such analyses often encompass aspects like colors, objects, scenes, and overall aesthetics~\cite{guntuku2019twitter, garimella2016social, Reece2017}.

\subsection{{Contextual Image Factors in Human Computer Interaction}}
Understanding the content and intrinsic characteristics of spontaneous images could be essential from a HCI standpoint. Contextual elements like environment, angle, color, and lighting play a significant role in how users interact with their smartphones. For example, research by \citet{ikematsu2020investigating}  indicates that people often prefer positions that require minimal movement when using their devices. This makes it valuable to examine factors such as the smartphone's angle and the background objects present during use. In addition, the ambient light during device interaction can act as a situational impairment, as noted by \citet{tigwell2018s} and \citet{ sarsenbayeva2017challenges}. For instance, the facial expressions and illumination on a user's face can vary greatly between bright outdoor sunlight and controlled indoor lighting conditions. The environment, whether indoor or outdoor, also affects color, which in turn can influence user psychology. \citet{valdez1994effects} conducted studies assessing the impact of color on emotions like pleasure, dominance, and arousal. Their findings suggest that colors like blue and purple are typically perceived as pleasant, while greenish hues tend to be more arousing. This raises the possibility that the dominant color in a user's surroundings might have a correlation with their facial features during smartphone interaction.

\subsection{{Smartphone Images in Controlled Settings for Mental Health}}

 Extracting facial features to assess mental health and emotions has received significant attention in computer vision, with applications spanning from education to healthcare~\cite{Mellouk2020}. Here, many studies have explored facial expressions, gaze patterns, and the overall composition of images to extract visual markers symptomatic of depression~\cite{Kong2022, Lee2022, Liu2022}. However, most of these studies are conducted in controlled environments or rely on participants deliberately capturing their images, which could inadvertently influence their emotional portrayal. For instance, ~\citet{Kong2022} captured photographs using a tablet in a standardized clinical setting. Participants were asked to sit before a white background, remove hats or glasses, and tie up long hair to expose their ears; the users looked straight ahead with relaxed expressions as instructed. Similarly, ~\citet{Liu2022} employed a multi-modal deep Convolutional Neural Network (CNN), considering both facial expressions and body movements. During psychotherapy sessions, they captured video using a 4K high-resolution camera in a controlled laboratory setting. Consequently, the participants' expressions and body movements were analyzed in a highly regulated context. Numerous other studies have similarly relied on advanced devices for image capture, used video recordings, or incorporated additional signals (such as movement, audio) within controlled environments~\cite{zhou2018visually, Guo2021, ramos2021evaluating, 7412257, joshi2022depression, Francese2022}. 

Our work aims to address these limitations by examining the feasibility of using spontaneously captured images from participants' smartphones, which offers a more natural and less intrusive method for predicting depression. As smartphones have become an integral part of modern life, they are an ideal tool for unobtrusive and widespread data collection. By utilizing smartphone cameras to capture participants' images, our approach eliminates the need for controlled environments or deliberate image-taking, thereby reducing the potential for biased emotional portrayals. Furthermore, the widespread availability of smartphones enables our method to reach a larger and more diverse population, ultimately promoting greater accessibility and inclusivity in mental health assessments.

\subsection{``In-the-wild" Smartphone Images for Mental Health}

Our study emphasizes the analysis of "in-the-wild" smartphone images, particularly those captured via front-facing cameras of smartphones. These images offer a direct window into an individual's emotions, expressions, and environment, thus enhancing the accuracy of mental health assessments. In contrast to social media content, these images remain relatively free from biases like social desirability and self-presentation, which often affect traditional tools. A limited number of past research have used "in-the-wild" smartphone images for mental health evaluation. For instance, ~\citet{10.1145/2800835.2804391} collected 5811 opportunistic photos in-the-wild from 37 students over ten weeks using their phone's front-facing camera. The study reported that depression scores significantly correlate with the students' facial expressions and activity. While ~\citet{10.1145/2800835.2804391} was the first to use in-the-wild images from front-facing phone cameras to study mental health on a non-clinical population of college students, the authors state that there was insignificant signal in the images to predict self-reported depression. MoodCapture is inspired by this original work, which was part of the StudentLife study ~\cite{wang2014studentlife} in 2013. Our progress is that a decade on from the StudentLife study, phone cameras have seen significant advancements, leading to substantial differences in their capabilities compared to those from ten years ago. For example, new phone cameras typically offer much higher resolution and more megapixels than those from a decade ago, resulting in sharper and more detailed face photos; advances in sensor technology and image processing have greatly improved low-light performance, resulting in today's phone cameras capturing better quality face photos in low-light conditions; optical image stabilization has become more common in smartphone cameras today, reducing the impact of shaky hands and resulting in smoother sharper photos, especially in low light; and finally front-facing cameras primarily designed for selfie shots have improved significantly in terms of resolution, image quality, auto-focus on the face. Other differences between ~\citet{10.1145/2800835.2804391} and our work are that we take advantage of massive advances presented by deep learning models and focus not on a non-clinical group but a clinical population.

Other studies have also leveraged front facing cameras in one way or another. ~\citet{Khamis2018} studied the visibility of the face and eye in 25,726 in-the-wild images of smartphone users and found that the full face is visible about 29\% of the time. The authors stated that their state-of-the-art face detection algorithm performed poorly against photos taken from front-facing cameras. Similarly, ~\citet{bace2020quantification} used in-the-wild images to study the visual attention and gaze of users. ~\citet{darvariu2020quantifying}, on the other hand, used in-the-wild images from rear-facing cameras. The authors developed a smartphone application that allows users to periodically log their emotional state together with
 pictures from their everyday lives. They collected 3,305 mood reports with photos from 22 participants. Authors report finding context-dependent associations between
 objects surrounding individuals and their self-reported emotional state. However, the genuine spontaneity of these captures and their potential for unbiased mental health evaluation remain relatively unexplored. Our contribution to this growing field pivots on the innovative use of genuinely spontaneous, in-the-wild facial images for depression detection. By employing a passive-sensing mobile application that seamlessly captures images without the subject's acute awareness, we negate the potential influence of self-awareness on emotional representation. This strategy bolsters the ecological validity of our data source, making it a robust tool for depression detection. 
\section{Methodology}
\label{sec:methodology}
In what follows, we discuss the design of our MoodCapture study, demographic information of the individuals that participated in the study and the ground-truth used for analysis.

\subsection{Study Design}
\begin{figure*}
     \centering
     \begin{subfigure}[b]{0.24\textwidth}
         \centering
         \includegraphics[width=0.8\textwidth]{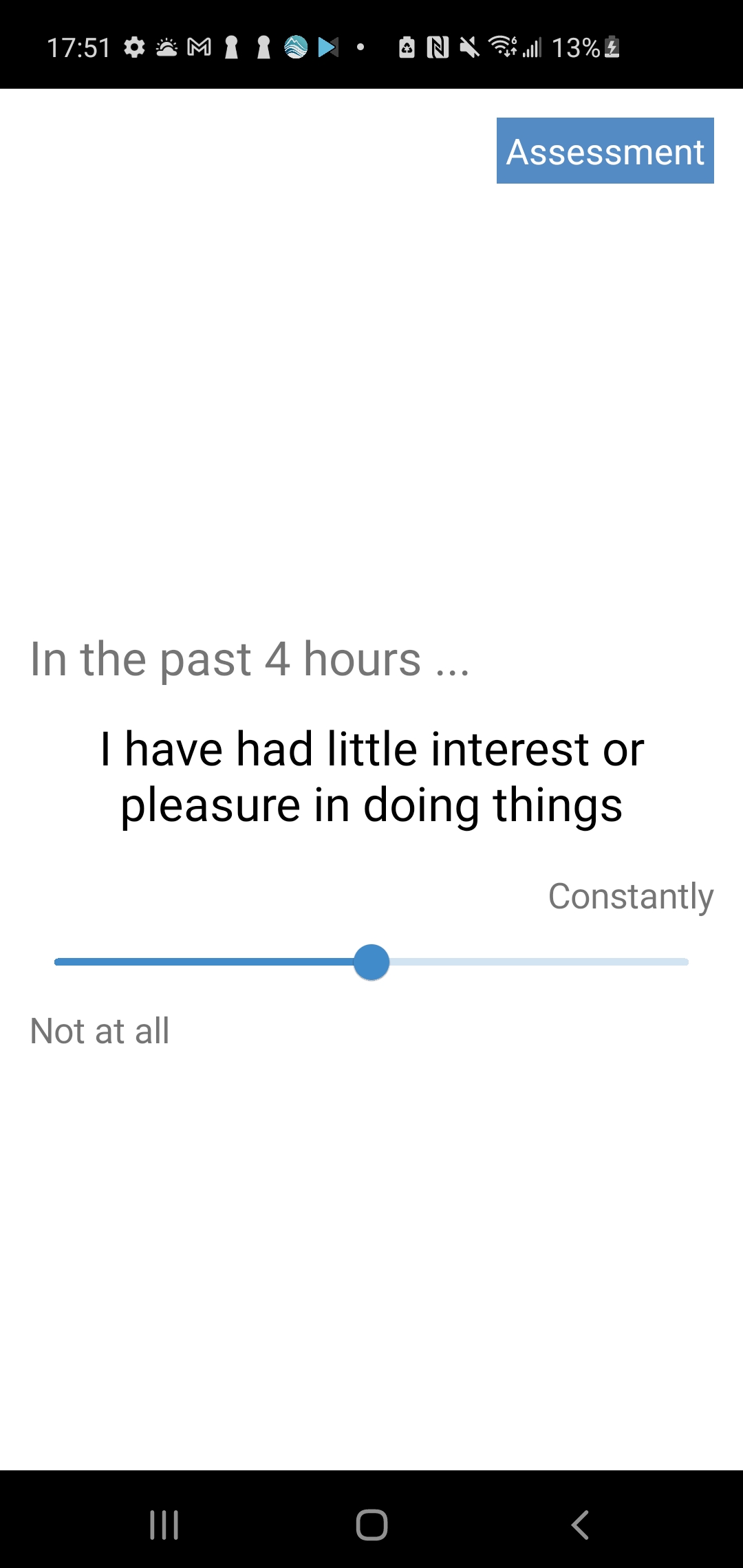}
         \vspace{0.5cm}
     \end{subfigure}
     \begin{subfigure}[b]{0.24\textwidth}
         \centering
         \colorbox{cyan}{\includegraphics[width=0.8\textwidth]{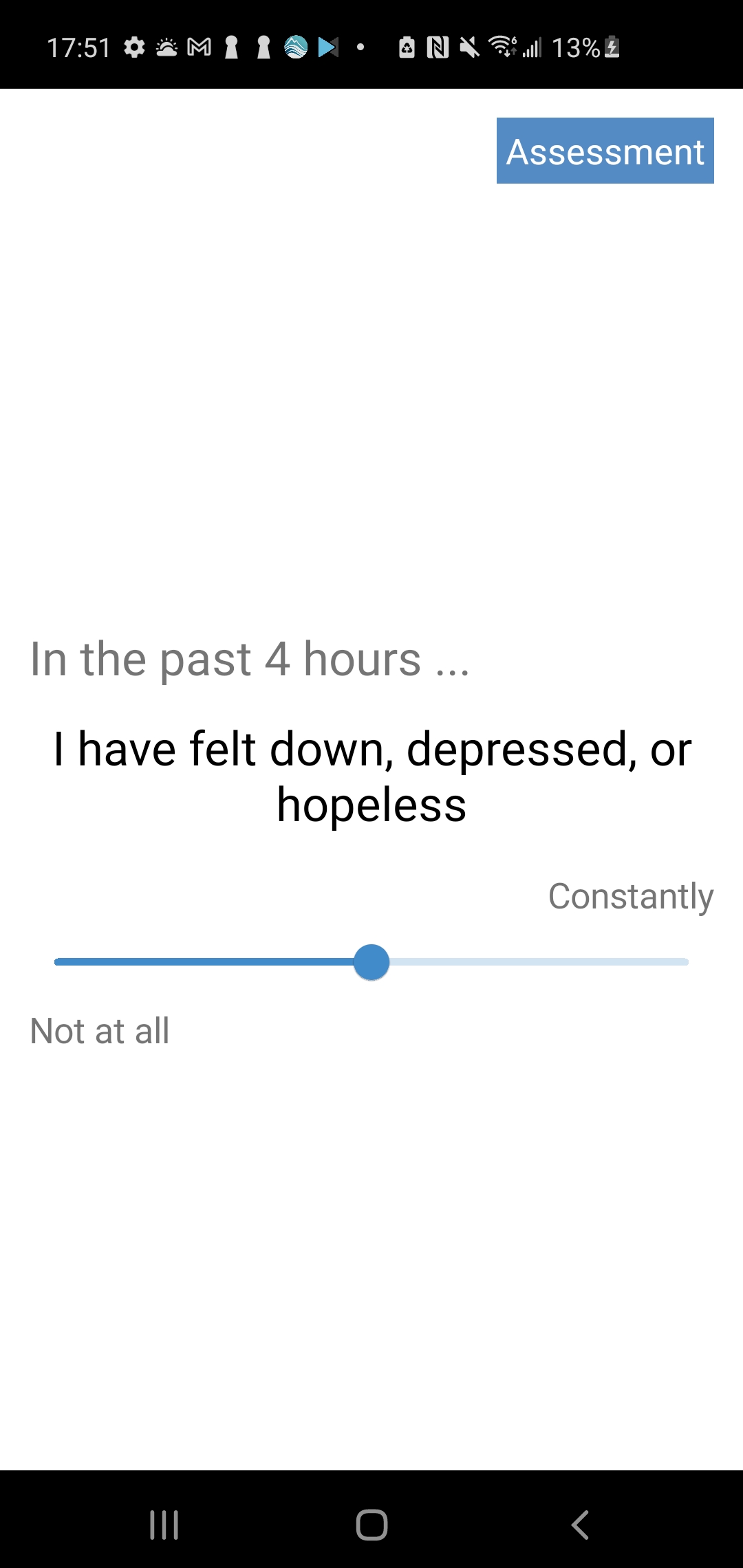}}
                  \vspace{0.5cm}
     \end{subfigure}
     \begin{subfigure}[b]{0.24\textwidth}
         \centering
         \includegraphics[width=0.8\textwidth]{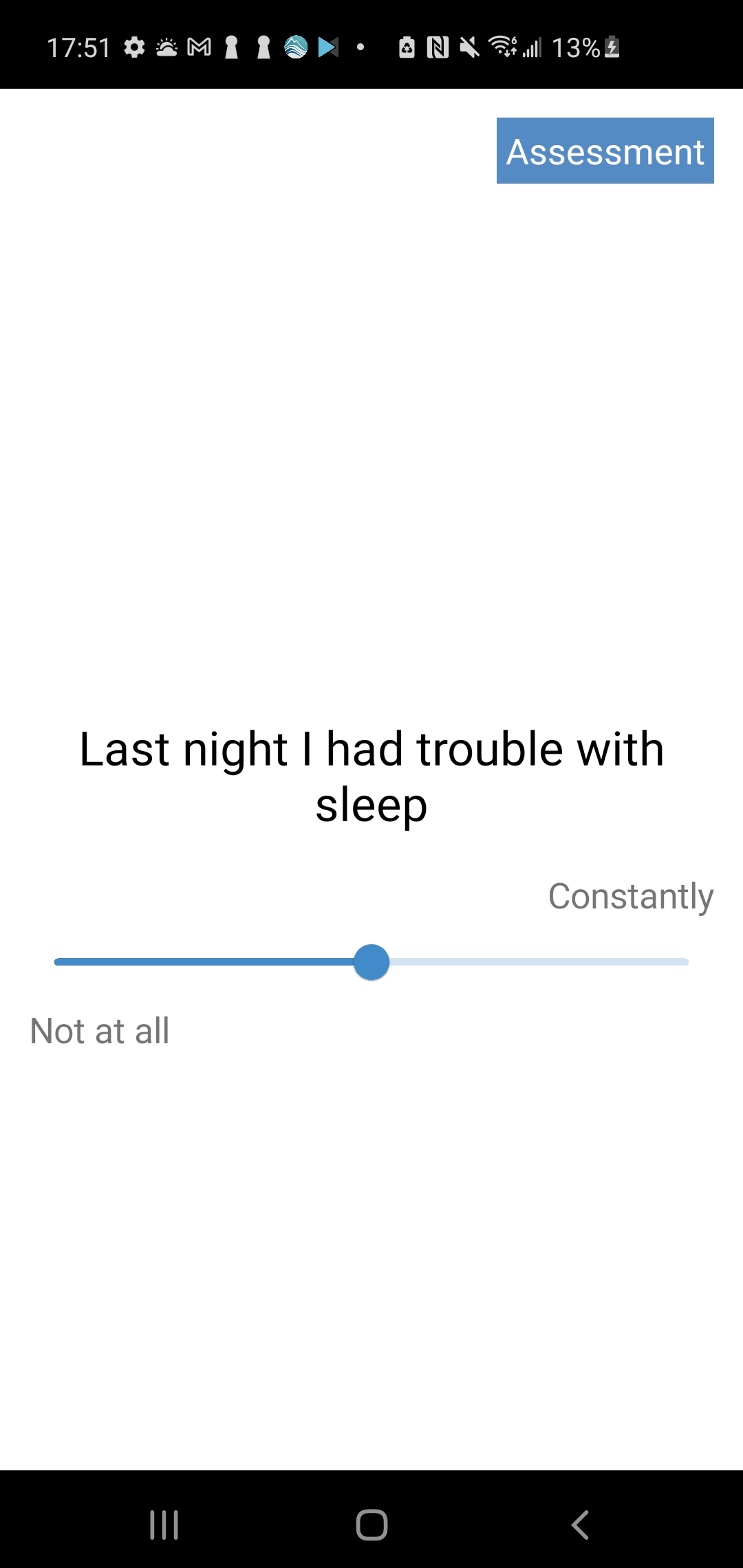}
                  \vspace{0.5cm}
     \end{subfigure}
    \begin{subfigure}[b]{0.24\textwidth}
         \centering
         \includegraphics[width=0.8\textwidth]{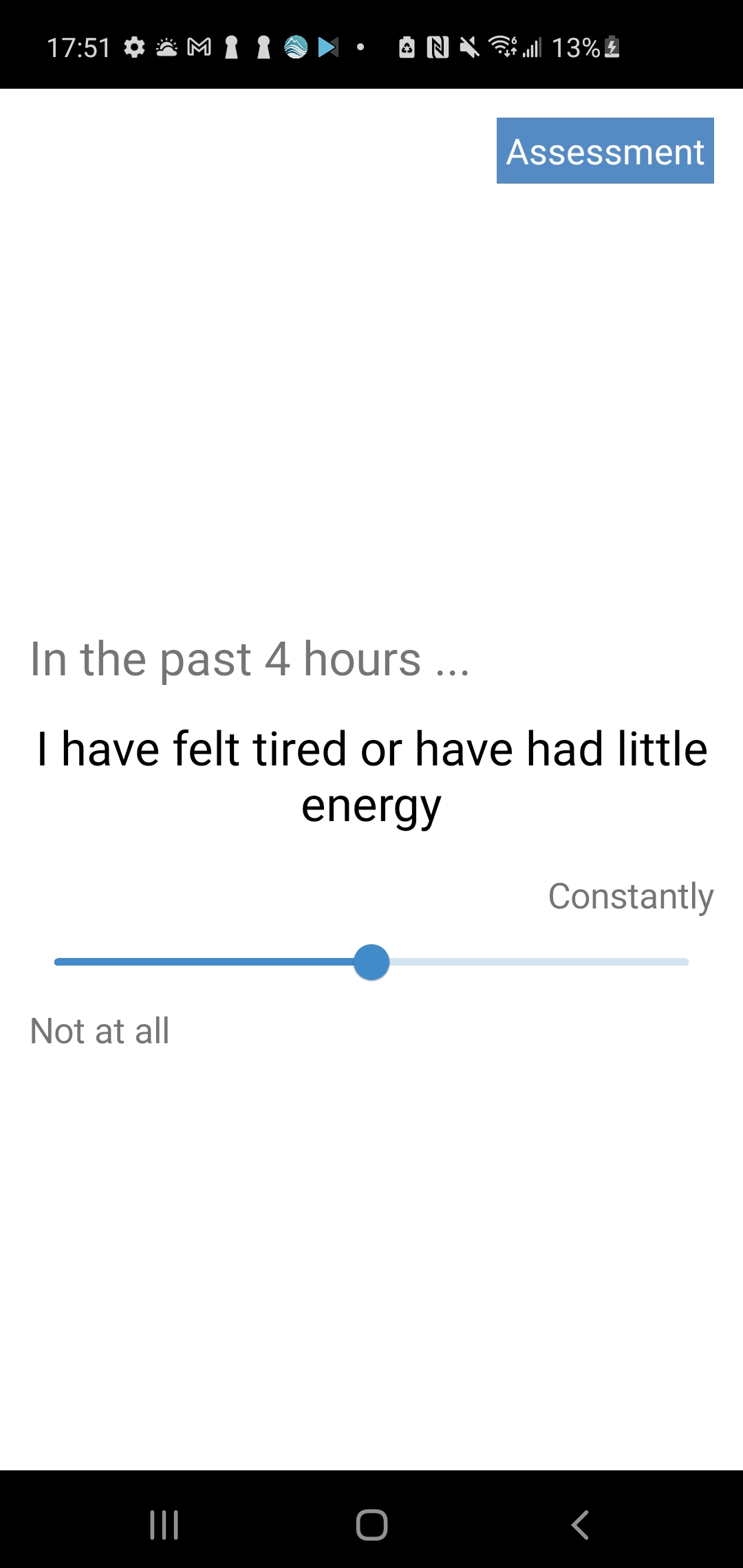}
                  \vspace{0.5cm}
     \end{subfigure}
    \begin{subfigure}[b]{0.24\textwidth}
         \centering
         \includegraphics[width=0.8\textwidth]{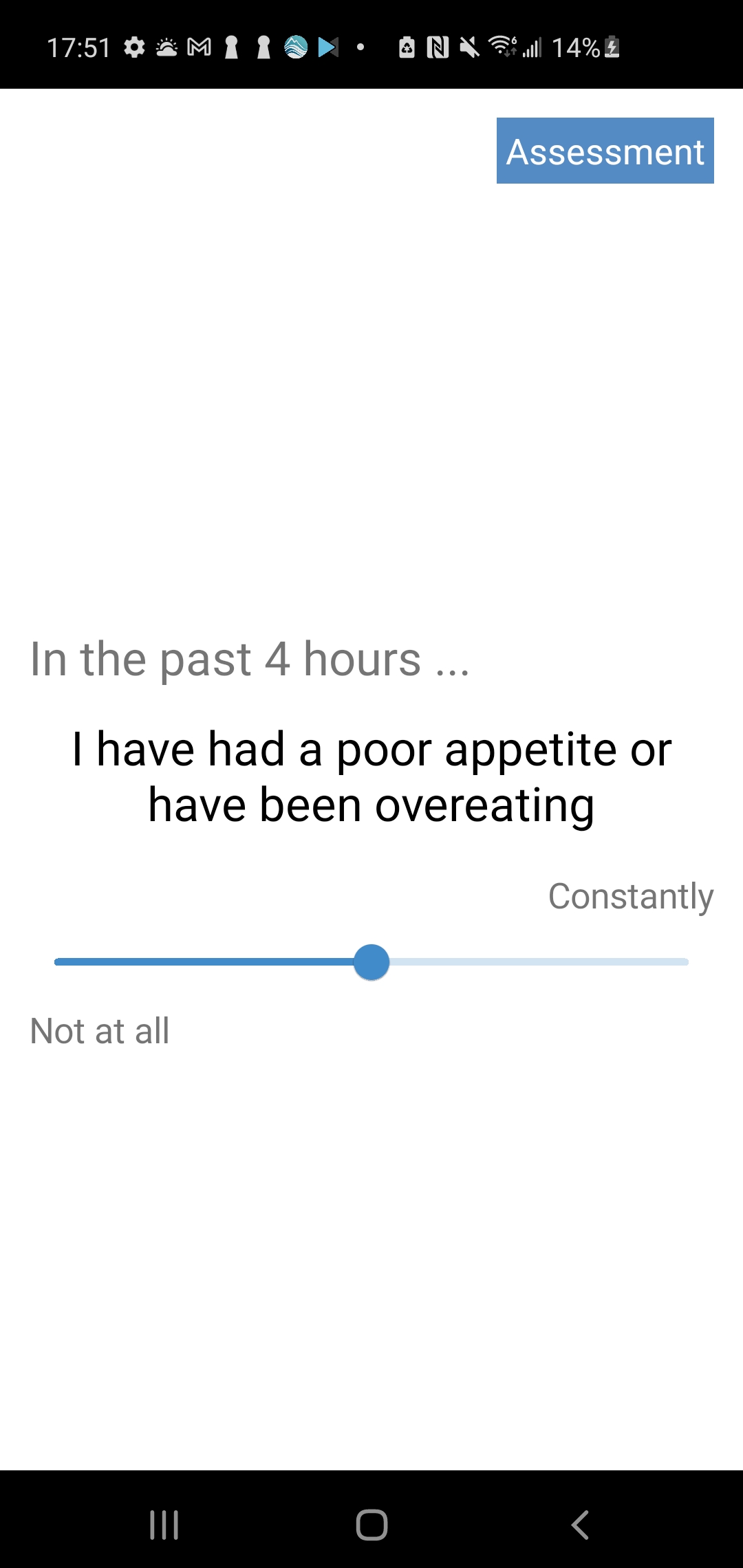}
     \end{subfigure}
         \begin{subfigure}[b]{0.24\textwidth}
         \centering
         \includegraphics[width=0.8\textwidth]{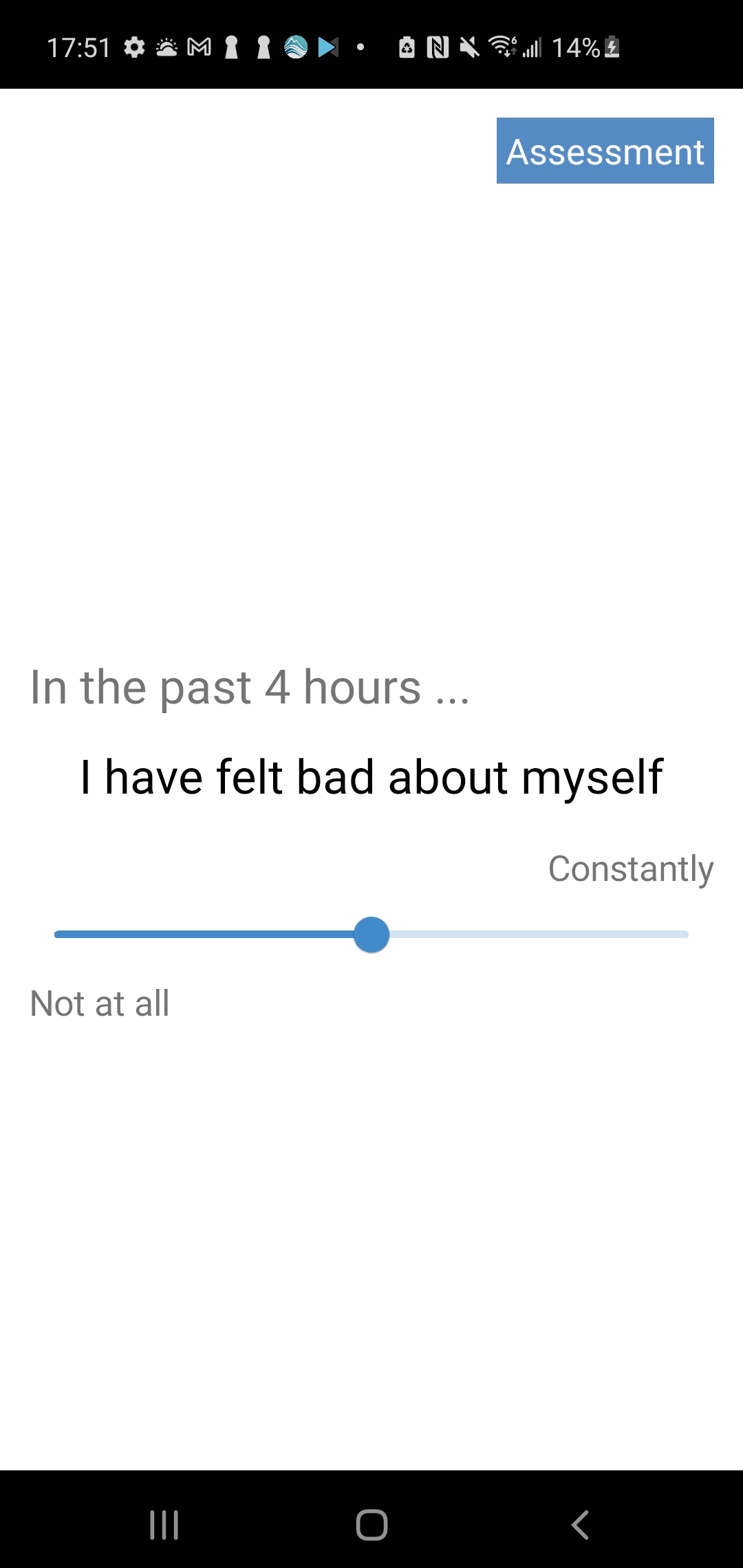}
     \end{subfigure}
         \begin{subfigure}[b]{0.24\textwidth}
         \centering
         \includegraphics[width=0.8\textwidth]{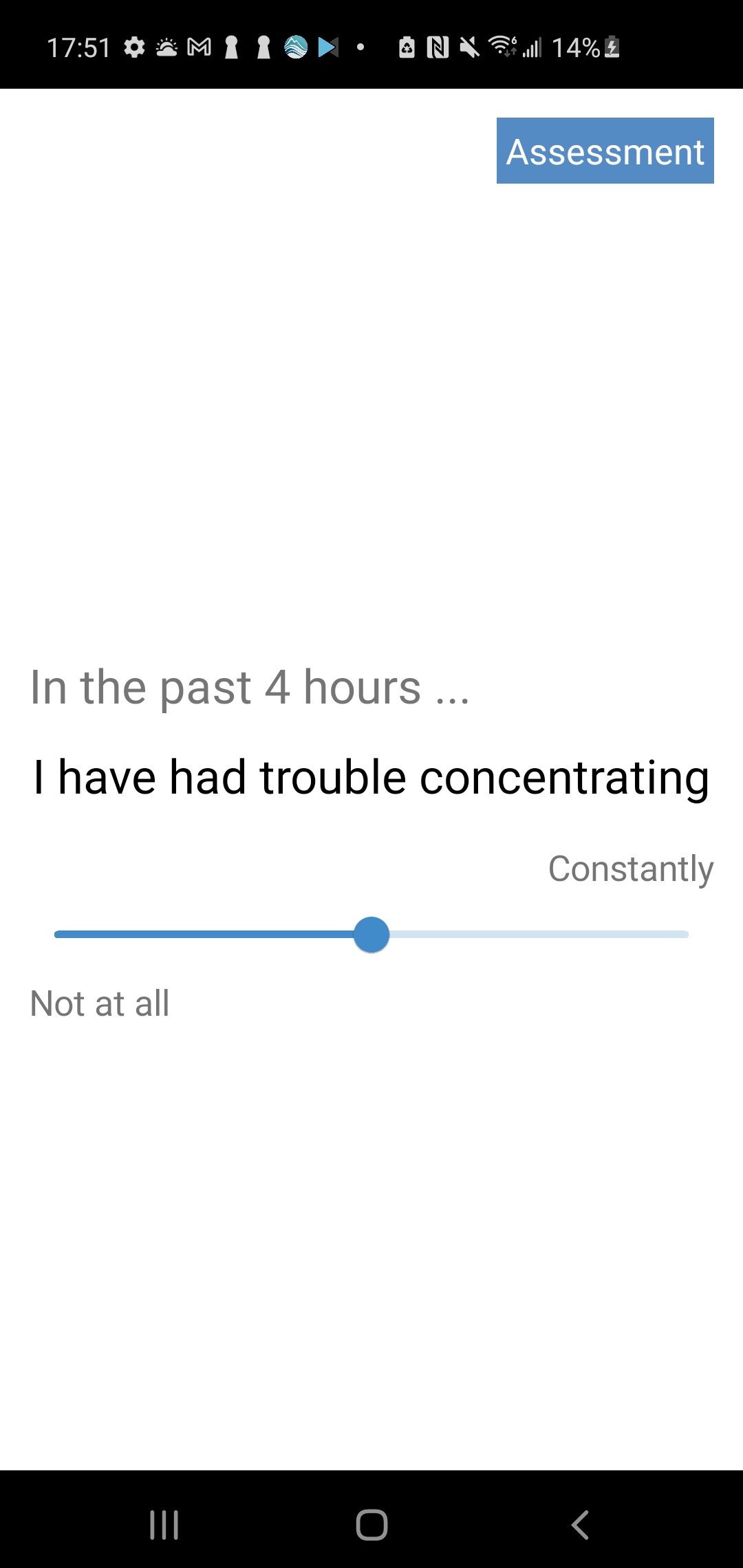}
     \end{subfigure}
         \begin{subfigure}[b]{0.24\textwidth}
         \centering
         \includegraphics[width=0.8\textwidth]{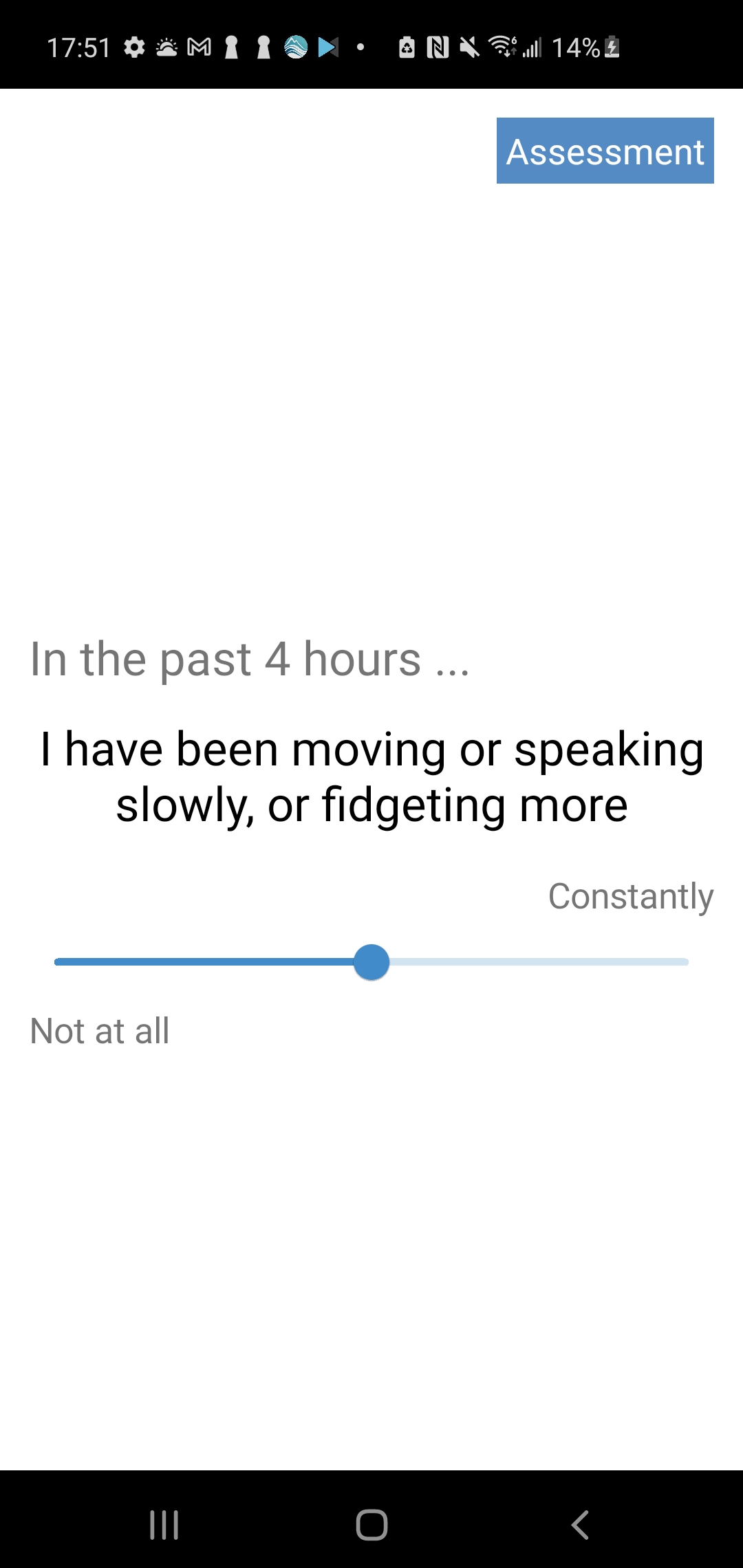}
     \end{subfigure}
        \caption{PHQ-8 application screens for each item:  Images are always captured while users respond to the PHQ-8 depression survey question (highlighted in cyan): ``I have felt down, depressed, or hopeless''. While users consent to have photos taken using the front-facing camera during the operation of the MoodCapture app they are not informed exactly when these photos are captured to promote in the moment naturalistic and authentic images.}
        \label{fig:phq8screens}
        \Description{Figure shows the screencap of the 8 questions as they are asked in our mobile application, with the second question where we take the picture of the participants highlighted.}
\end{figure*}

We recruited 181 participants from across the United States using targeted online advertisements on Google and Facebook. Each participant underwent a clinician-administered Structured Clinical Interview for DSM-5 (SCID), and only those diagnosed with Major Depressive Disorder (MDD), without bipolar disorder, active suicidality, or psychosis, were eligible for the study. Upon qualification, participants installed our Android-based mobile sensing app on their devices, which gathered Ecological Momentary Assessments (EMA) during the 90-day study period. The app prompted participants to complete a brief Patient Health Questionnaire-8 (PHQ-8)~\cite{kroenke2009phq} (see Table 6) survey about their depressive symptoms three times daily (morning, afternoon, and evening). As participants answered their daily surveys, the app was designed to discreetly capture a burst of up to 5 images using the front-facing camera. Specifically, images were taken when participants responded to the PHQ-8 item: \textit{``I have felt down, depressed, or hopeless.''} (see Fig. \ref{fig:phq8screens}). We chose this question as we believed it would best capture participants' genuine emotions related to depression. The PHQ-8 is a validated inventory for measuring depression. For further information about the survey, please refer to the Ground Truth section.

During the onboarding process, we informed participants about the image capture procedure and emphasized that sharing their photos was optional. Upon launching the mobile app for the first time, participants were asked, \textit{``To help us better understand your depressive symptoms, we would like to take a few photos in the background that capture your facial expressions while you fill out questionnaires. Do you give us permission to do this?''} Participants could respond with either "Yes" or "No." If they agreed to share their photos, the app captured images as they answered the EMA. If they opted not to share their photos, no images were captured. The image capture process was designed to be unobtrusive, with only a green dot at the top of the Android status bar/screen indicating camera usage -- which users' may or may not have observed. Participants did not see their face or receive any other indication that photos were being taken. This discreet image capture process ensured a seamless user experience without interrupting or obstructing the EMA flow. As stated earlier; while participants consented to have photos taken using the front-facing camera during the operation of the MoodCapture app in the study they were not informed exactly when these photos were captured, thus promoting in-the-moment naturalistic and authentic capture of users' faces and surroundings.

Participants were compensated \$1 for each completed EMA, with an additional \$50 bonus for achieving a completion rate of 90\% or higher during the study period. Compensation was not dependent on sharing photos; participants were compensated regardless of their photo consent. The study was approved by Dartmouth College's Internal Review Board (IRB). Our analysis and predictive modeling focuses on 177 out of the 181 participants who provided consent for their photos to be captured. We collected 125,335 images from these participants, excluding 15,063 photos that were either too blurry, contained no faces, featured children, or contained nudity.

\subsection{Demographics}
The majority of participants in our study identified as female (86.4\%, N=153) followed by male (9.6\%, N=17) and non-binary (2.8\%, N=5). In terms of race, 83.6\% (N=148) are White, 2.8\% (N=5) are Asians, 4.5\% (N=8) are Black or African American, 0.5\% (N=1) are American Indian/Alaska Native and 6.7\% (N=12) belong to more than one race. See Table~\ref{tab:demographics} for the detailed breakdown.

\begin{table}[ht!]
\caption{{Demographics, smartphones, and image composition in our study.}}
\begin{tabular}{@{}lllll@{}}
\textbf{Category}                                                                                                                                            & \textbf{Count}   & \textbf{Percentage}\\ \bottomrule
\multicolumn{3}{l}{\cellcolor[HTML]{EFEFEF}\textit{Sex}}                                                                       \\
Female & 153 &86.4\%    \\
Male   & 17 & 9.6\%    \\
Non-binary &5 &2.8\%\\
Other (prefer to self-describe) &2 &1.1\%\\ 
\multicolumn{3}{l}{\cellcolor[HTML]{EFEFEF}\textit{Race}}    \\
White & 148 &83.6\%  \\
Asian & 5  &2.8\%  \\
Black or African American & 8 &4.5\%   \\
American Indian/Alaska Native & 1 &0.5\%   \\                   
More than one race & 12 &6.7\%   \\ 
Other (prefer to self-describe) &3 &1.6\% \\
\multicolumn{3}{l}{\cellcolor[HTML]{EFEFEF}\textit{Smartphones}} \\
Samsung &107 &60.4\% \\
Google &36 &20.3\% \\
Motorola &19 &10.7\% \\
Other &15 &8.4\%\\
\multicolumn{3}{l}{\cellcolor[HTML]{EFEFEF}\textit{Image Resolution}} \\
3648x2736 &57 &32.2\% \\
3264x2448 &52 &29.3\% \\
2640x1980 &16 &9.0\% \\
Other &52 &29.3\% \\
\bottomrule
\end{tabular}
  \label{tab:demographics}
\end{table}

\subsection{Ground Truth}
\label{sec:groundtruth}

Our study is designed to account for the wide variability in MDD symptoms. In particular, MDD can manifest in over 1000 distinct symptom combinations across individuals, with significant within-day variations \cite{cramer2016major, fried2015depression, ebrahimi2021within, fried2022revisiting}. However, existing diagnostic methods face several limitations. Firstly, SCIDs are not effective in capturing moment-to-moment fluctuations in depression symptoms. Secondly, the Likert scale used in depression screening tools like the PHQ-8, which typically offers a limited response range from 0-3, forces respondents to fit their experiences into pre-set categories. This can lead to central tendency bias and a lack of detailed responses for complex mental states. To overcome these challenges and better capture intra-individual variation, our clinical team modified the PHQ-8 scale to a more nuanced continuous scale ranging from 0-100 (see Figure~\ref{fig:phq8screens}). The practice of re-scaling psychometric scales is not uncommon and
has been applied to the PHQ in various past studies~\cite{Majethia2022, Nguyen2021, Gumus2023, Lum2016}. A standard PHQ-8 score of 10 or higher (out of 24) signifies major depression \cite{phq2019}. In our continuous scale, a score exceeding 334 indicates depression (i.e., 10/24 times 800). To provide holistic analysis, we complement our binary classification models with regression models that predict raw PHQ-8 scores. Note that the PHQ is versatile, serving both as a screening tool for depression and as a means to monitor clinical symptom changes~\cite{Kroenke2010-qi}.

\begin{figure}[h]
    \centering
    \begin{subfigure}{0.48\textwidth}
      \centering
    \includegraphics[width=.9\linewidth]{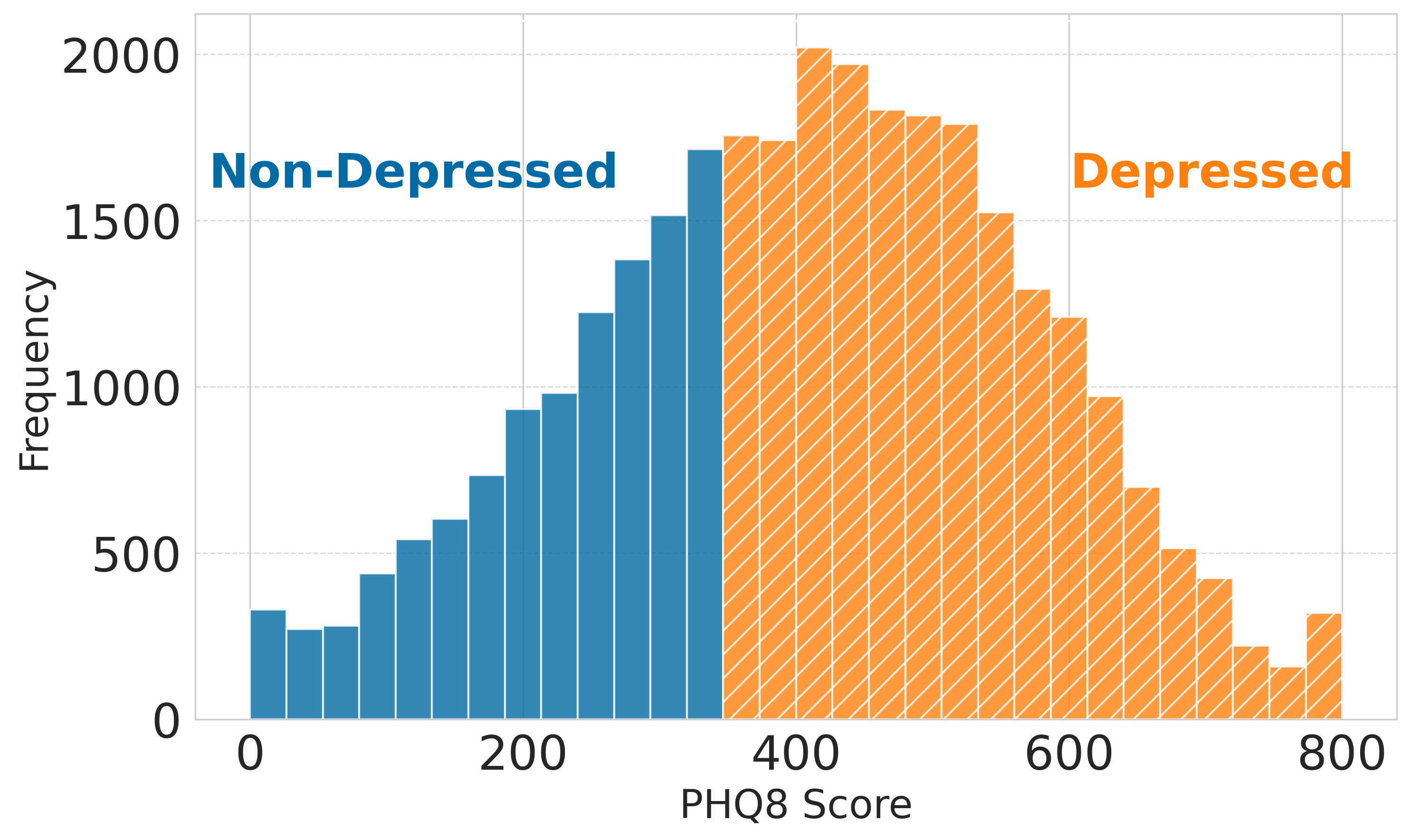}
      \caption{Distribution of PHQ-8 Scores}
      \label{fig:gtdistrib}
    \end{subfigure}
     \begin{subfigure}{0.48\textwidth}
      \centering
    \includegraphics[width=.9\linewidth]{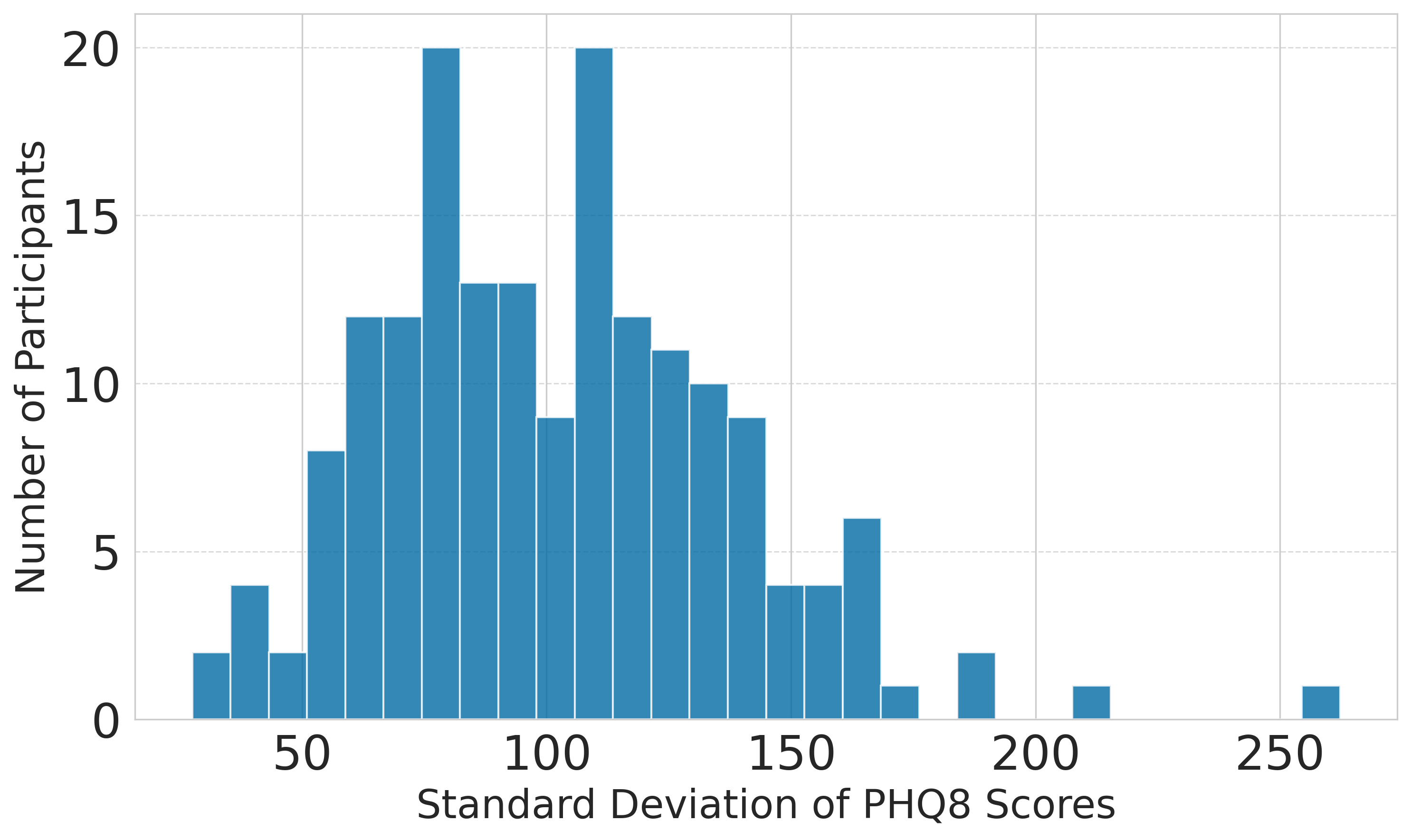}
      \caption{Intra-individual variability}
     \label{fig:stdfigure}
    \end{subfigure}
    \caption{PHQ-8 score statistics: Figure (a) depicts the distribution of the PHQ-8 score reported by the participant and the corresponding label (i.e., Depression or No Depression). Figure (b) shows the variability of PHQ-8 scores among participants over the duration of the study (Cronbach's $\pmb{\alpha=0.85}$).}
 \Description{The top figure shows the distribution of the EMA self-reports we obtain and we separate them into two groups: Non-Depressed and Depressed based on whether the self-reported PHQ-8 values is under or over the given threshold. There is almost an even split in the distribution, with depressed self-reports slightly more than non-depressed. Bottom figure plots the standard deviation of the obtained self-reports. It is mostly less than 150.}
\end{figure}

To enhance the reliability and accuracy of the EMA responses, we employed a validation technique wherein the app randomly reversed one question in each PHQ-8 survey (thus adding an additional item), ensuring that participants are attentive. We then compared the responses to the original and reversed questions; if there is a significant discrepancy, the response is excluded from our analysis. After applying this filtering process, we obtain a refined dataset comprising 31,215 EMAs. Since we captured a burst of images with each EMA response, we amassed 110,272 images in total. As depicted in Figure ~\ref{fig:gtdistrib}, we divided our dataset into two groups: depressed (74,347 images, N=175) and non-depressed (35,925 images, N=156). On average, participants submitted 176 EMAs (stdev = 78) and 623 images (stddev = 278) per participant during the study period. It is crucial to note that all participants recruited for this study had major depressive disorder. Consequently, they reported being below the cut-off threshold on some days and above it on others. However, 19 participants consistently reported depression throughout the study.

Figure~\ref{fig:stdfigure} shows the variability of PHQ-8 scores among participants i.e., intra-individual variability. It provides insight into the fluctuations in a participant's scores over time. On average, participants' scores varied around their own mean by approximately 101.92 points, with the variability ranging widely from a standard deviation of 27.56 points to as high as 262.24 points. This suggests that some participants had relatively stable scores over time, while others exhibited more pronounced fluctuations. Moreover, we measured the internal consistency of the PHQ-8 items, obtaining a Cronbach's $\alpha=0.85$. This demonstrates good reliability and validity of our measures.

\subsection{Image Characteristics}
\label{subsec:image_characteristics}
We gather in-the-wild images captured by participants using a diverse range of smartphones with varied configurations and camera placements. Predominantly, participants use Samsung, Google, and Motorola devices, and the images captured from these devices had resolutions ranging from 1920x1080 to 4656x3488 (see Table \ref{tab:demographics}). Our naturalistic approach at capturing image ensures ecological validity and represents users' natural behavior while engaging with their devices in different environments. To examine the characteristics of these images, we analyze factors such as phone angle, dominant color, lighting condition, photo location, and background elements present in the photos. The in-the-wild smartphone images offer a unique glimpse into the multitude of ways users interact with their devices and surroundings. However, extracting meaningful insights from these images demands a refined approach that acknowledges the diverse contexts in which they are captured. To achieve this, we utilize the BLIP \cite{li2022blip} visual question answering (VQA) model, an advanced AI tool specifically designed for image analysis and answering questions about image content and context. BLIP is recognized as a state-of-the-art method for visual question answering tasks. Furthermore, the VQA analysis contextualizes our predictive modeling in the following ways. First, it can elucidate the raw image content, which is the input for our deep learning models. Second, as our ML models use handcrafted features from the face, it differentiates the performance obtained by considering background in addition to face versus only face. In summary, our motivation is to harness VQA to interpret both explicit and implicit image content. Consequently, enabling a more holistic approach to image analysis, where both the central subjects and their surrounding context contribute to the predictive insights. Importantly, as we cannot display images to protect participant privacy, the VQA provides some level of interpretation. With the help of the VQA model, we explore the following characteristics:

\para{Image Angle:} By inquiring about the image angle, we gain an understanding of user interaction dynamics with their devices. Varying angles, such as high or low, offer insights into users' physical engagement with their smartphones. High, low, or level angle refers to the perspective from which an image is captured or taken with respect to the subject in the frame. A low angle shot refers to the subject looking down at their phone, whereas a high angle shot refers to the user looking up at their phone. A level angle shot is taken from the same height as the subject, capturing it at eye level. We asked the VQA: \textit{``Is the image taken from a high, low, or level angle?''}.

\para{Dominant Colors:} Colors are crucial for establishing the context of an image. To identify dominant colors in the images and understand the users' environments, we asked the VQA: \textit{``What is the dominant color of the image?''}.

\para{Lighting Condition:} Lighting conditions in an image reveal important information about the user's ambient environment. Using the VQA model, we classified images based on their lighting as well-lit, dimly lit, or poorly lit. We asked the VQA: \textit{``Is the image well-lit, dimly lit, or poorly lit?''}.

\para{Photo Location:} The location context (indoors or outdoors) can significantly influence user-device interactions. We determined the location context of images with the help of the VQA model by asking: \textit{``Is the photo taken indoors or outdoors?''}.

\para{Background Objects:} Identifying specific objects in the background can provide valuable information about the user's context and activities. We queried the VQA model about the background objects to recognize and categorize various elements within the images. We asked the VQA: \textit{``What are the background objects in the photo?''}.

\para{Number of People in the Image:} In order to evaluate the social context of the images, we employed the VQA model to determine the number of people present in each image. This information provides insight into users' social interactions and their surroundings during device usage. We asked: \textit{``How many people are in the image?''}.\\
\newline
By leveraging the BLIP VQA model, we are able to extract structured insights about the content and context of in-the-wild images, enhancing our understanding of user behavior and interaction with their devices in diverse settings. Importantly, two expert annotators manually annotated 1500 unique images corresponding to individual EMAs. To clarify any ambiguities, we provided them with specific instructions. They determined the image angle based on eye level with the phone. `Dominant color' refers to the most prominent color in the overall image. For lighting conditions, `well-lit' represents the best lighting condition, while `poorly lit' indicates the worst. After completing the manual annotation, we calculated the average accuracy between the two annotators and the inter-rater agreement using Cohen's kappa ($\kappa$). These results (see Table \ref{tbl:communication_style}) indicate substantial agreement between the annotators and alignment with VQA responses, indicating high reliability, consistency and
accuracy.

\begin{table}[ht!]
\smaller
\caption{Description of OpenFace features \cite{baltrusaitis2018openface}.}
\begin{tabular}{@{}llp{0.55\columnwidth}@{}}
\toprule
\textbf{Feature Set} &\textbf{Features} &\textbf{Description} \\
\midrule
FAU~\cite{baltruvsaitis2015cross} & 35 & FAU uses the facial action coding system to describe anatomically possible expressions resulting from muscle activations. It indicates the presence and intensity of an expression\footnote{https://www.cs.cmu.edu/~face/facs.htm}. \\ \midrule
Gaze \cite{wood2015rendering} &8 &Direction vector in eye gaze direction as measured from the eye location, consisting of location and angle. This feature is computed after landmark detection. \\ \midrule
Eye Landmarks &280 & Contains 2D and 3D landmarks describing various positions of the eye. \\ \midrule
Head Pose &6 & Describes translation in millimeters with respect to camera centre and rotation in radians around x, y, and z axes. \\ \midrule
Rigidity Parameters &40 & These parameters are divided into rigid and non-rigid shape parameters. Rigid shape parameters describe the face's positioning within an image that includes aspects like scale, rotation, and movement. Conversely, the non-rigid shape parameters focus on the variations in facial appearance caused by individual characteristics or expressions, such as variations in facial width or height, smiles, blinking, and other facial expressions.\\ \midrule
2D Landmarks \cite{baltrusaitis2013constrained} &136 & These are x and y axes locations of different face landmarks in the image. These landmarks refer to specific locations in the face. For example, a point in the right eye is represented as landmark number 38, while points in the lips are represented using numbers 49-68. All landmark numbers are described in~\cite{sagonas2013300, sagonas2016300}. \\ \midrule
3D Landmarks \cite{baltrusaitis2013constrained} &204 &These are x, y, and z axes locations of different face landmarks in the image. The landmark numbers are identical to 2D landmarks, however, they are represented using three coordinates. \\
\bottomrule
\end{tabular}
\label{tab:openface_features}
\end{table}

\subsection{Depression Classification and Regression}  
\label{subsec:depression_detection}  
In this study, we aim to accurately identify depression from facial images by utilizing both machine learning and deep learning techniques. In particular, we build binary classification models to classify a face image as depressed or not depressed, and a regression model to predict the raw PHQ-8 score (see Section \ref{sec:groundtruth}).  
   
\subsubsection{Machine Learning} \label{subsubsec:machine_learning}  
To facilitate machine learning approaches, we extract 711 (709 trainable) facial features using OpenFace \cite{baltrusaitis2018openface}, a well-validated feature set for depression detection that has been employed in a variety of studies \cite{gong2017topic, saggu2022depressnet, pampouchidou2017facial}. The extracted features consist of 2D and 3D facial landmarks, head pose, eye gaze, facial expressions represented by facial action units (FAU), and rigid and non-rigid shape parameters (see Table \ref{tab:openface_features}). Before training, we apply feature selection using only the training set in two distinct ways. First, we compute the mutual information (MI) metric, selecting the most independent features indicated by smaller MI values. In our analysis, we choose the top 25\%, 50\%, or 100\% of the features. Second, we conduct an ablation study to gain valuable insights into the effectiveness of different hand-crafted features, thus inferring the best performing feature set. An ablation study is a systematic experimental procedure in which certain features are systematically removed or ``ablated'' to analyze their individual impact on the overall model performance.
 We use a Logistic Regression \cite{hosmer2013applied} model for classification and an ElasticNet \cite{zou2005regularization} for our regression task, whereas a Random Forest (RF) \cite{breiman2001random} is used for both tasks. Statistical approaches such as regression and a bagging-based decision tree can provide different modeling insights. The \textit{baseline} model is a RF trained using the participant's gender, age, and time spent on EMA.
   
\subsubsection{Deep Learning} \label{subsubsec:deep_learning}  
Deep learning models are capable of learning useful features directly from raw images. Pre-trained computer vision models trained on large-scale datasets can capture image features that are transferable to other domains. As a result, we examine the performance of various EfficientNet \cite{tan2019efficientnet} and InceptionResNetv3 variants, which were previously trained on the ImageNet and VGGFace2 datasets, respectively.  Upon observing that the EfficientNet B0 (EffNet) model provided the best performance while other models were underfitting our dataset, we decided to further fine-tune EffNet for depression prediction. We implement EffNet using the PyTorch framework, freezing all layers during the training process except for blocks 6 and 7. The classification and regression models are fine-tuned using binary cross-entropy and mean absolute error loss functions, respectively. For optimization, we use the Adam optimizer (with a learning rate of 0.0001) with a batch size of 256 trained for 50 epochs. This fine-tuning process allows the model to learn and adapt to the specific characteristics of our depression detection dataset, potentially improving its performance and generalizability.

\subsubsection{{Evaluation}} \label{subsubsec:evaluation}  
To effectively evaluate our models, we adopt a 5-fold leave-subject-out cross-validation approach. This method ensures that all images associated with a single participant are exclusively used for training, validation, or testing the model but not mixed among the subsets. Furthermore, we use nested cross-validation on our training data for hyper-parameter tuning. The subject-independent splits and cross-validation ensure our results are more robust than those of a single train-test strategy. We evaluate classification performance using balanced accuracy (Equation \ref{eqn:ba}) and Matthew’s Correlation Coefficient (MCC) (Equation \ref{eqn:mcc}), whereas regression performance is evaluated using MAE (Equation \ref{eqn:mae}). We chose these metrics as they provide a comprehensive assessment. For example, MCC summarizes all four values in the confusion matrix, whereas balanced accuracy emphasizes both true positive and true negative detection. In fact, MCC is preferred over F1 score in many binary classification problems \cite{chicco2020advantages}.

{\begin{equation}\label{eqn:ba}
    \text{Balanced Accuracy} = \frac{\frac{TP}{TP+FN} + \frac{TN}{TN+FP}}{2}
\end{equation}}

{\begin{equation}\label{eqn:mcc}
    MCC = \frac{TP \times TN - FP \times FN}{\sqrt{(TP+FP)(TP+FN)(TN+FP)(TN+FN)}}
\end{equation}}
where $TP, TN, FP,$ and $FN$ are true positives, true negatives, false positives, and false negatives, respectively. Note that higher balanced accuracy and MCC values indicate better performance. MCC ranges from -1 to +1, where +1, 0 and -1 indicate perfect classification, random coin-toss classification, and perfect mis-classification, respectively. The regression models are evaluated using MAE defined as:
{\begin{equation}\label{eqn:mae}
    MAE = \frac{1}{N} \sum_{i=1}^{N}|y_i - \hat{y}_i|
\end{equation}}
where $N$ is the number of samples. $y_i$ and $\hat{y}_i$ are true and predicted PHQ-8 scores, respectively. Note that lower MAE values indicate better performance.

\section{Results}
\label{sec:results}
\textcolor{black}{In this section, we present the outcomes of our analysis, which includes an examination of image characteristics, an evaluation of the predictive capabilities of our machine learning models, and an ablation study. In addition, we identify crucial features integral to our models' performance and explore potential biases within these models.}
\subsection{Image Characteristics}
\label{sec:image_characteristics_result}
\begin{figure}[ht!]
    \centering
\includegraphics[width=0.47\textwidth]{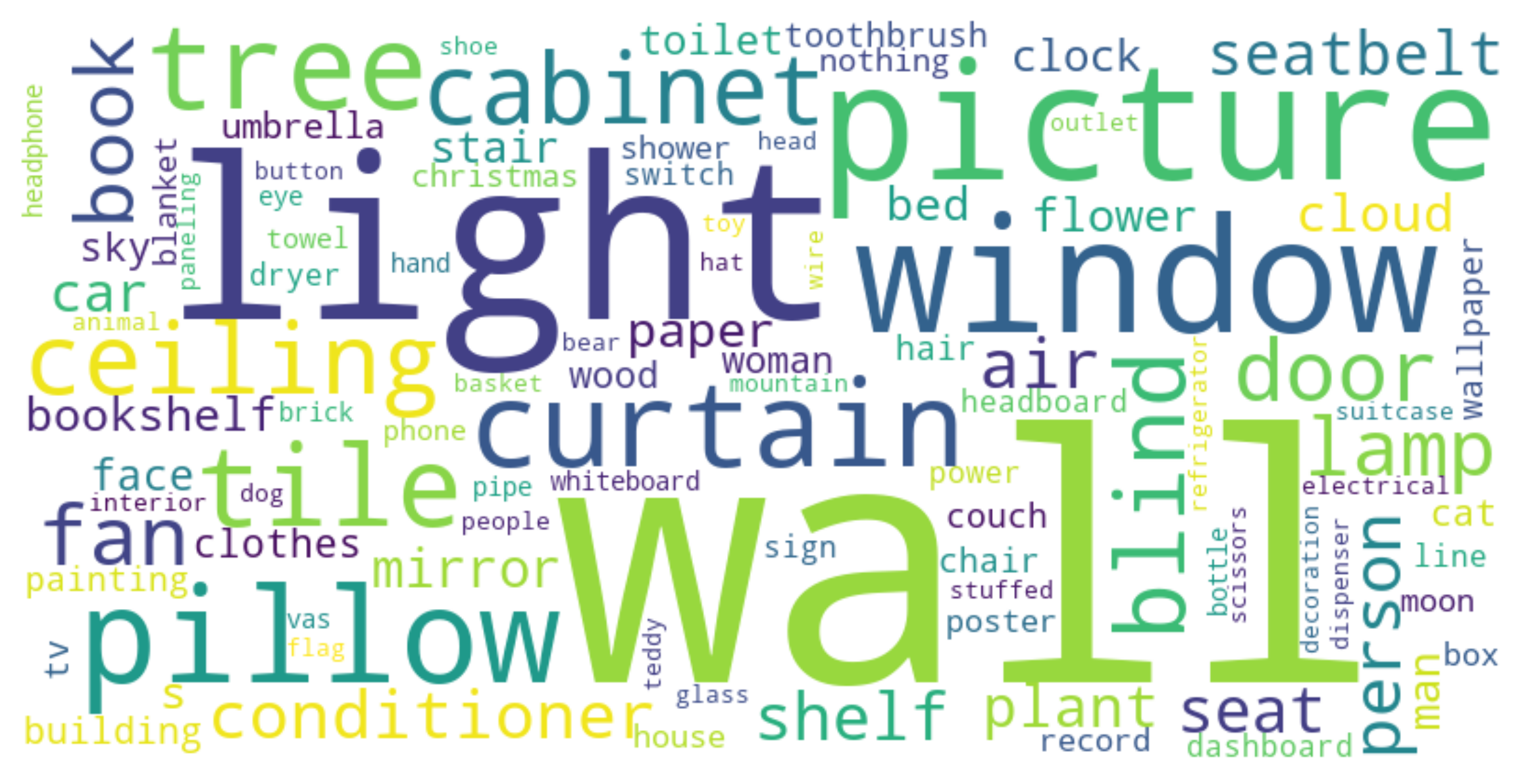}
    \caption{{Background objects: Word Cloud showing the range of objects detected in the background of the images captured. (Acc=91.72 ; $\pmb{\kappa}$=0.70)}}
    \label{fig:backgroundobjects}
          \Description{Figure shows a wordcloud that lists the most common background objects detected by our visual question answering model from the participants' images. Some of the frequent objects identified include wall, light, curtain, window, picture, pillow, cabinet, ceiling, tile, and book.}
\end{figure}

Our analysis using the VQA model reveal many insights into different features of real-world smartphone images. These images serve as glimpses into user interactions and surroundings. {From Table \ref{tbl:communication_style} and Figure \ref{fig:backgroundobjects}, we notice that the VQA obtained good accuracy ranging from 89\% for lighting conditions to 97\% for number of people in the image. Furthermore, $\kappa$ is greater than 0.70 for all questions indicating substantial inter-rater agreement.} In terms of capture angle, the images predominantly favored a low angle, with approximately 96.08\% falling into this category. Conversely, a mere 3.92\% were captured from a high angle, suggesting a specific user posture or device interaction habit in the majority of instances. Dissecting the dominant colors present, we found that `white' emerged as the prevailing color, characterizing roughly 67.51\% of the dataset. Other noticeable colors included `black' at 8.70\%, while a combined representation of `brown', `blue', `gray', and `yellow' accounted for approximately 18\%. A diverse array of other hues constituted the remaining 5.75\%, emphasizing the richness of user environments. Closer analysis  during the annotation process revealed that the images' dominant white color mainly reflects environmental elements like white walls and ceilings, not participants' skin tones. \textcolor{black}{Importantly, we noticed most images consisted of partial face images, an observation commonly found in other similar studies \cite{Khamis2018}. Hence, the dominant color is influenced by background objects.} This is evidenced in Figure~\ref{fig:backgroundobjects}, where walls, ceilings, tiles and lights are frequently identified as background objects, ensuring our analysis focuses on environmental, not physiological, aspects. The lighting conditions under which these images were taken were also revealing. 
A vast majority (80.57\%) were captured under well-lit conditions, indicating optimal settings for smartphone interaction. The dimly lit and poorly lit categories followed with 10.35\% and 9.08\%, respectively, showcasing the varied ambient conditions in which users interact with their devices. Furthermore, in terms of photo location, an impressive 95.08\% of the images were taken indoors, signifying the primary environment for user-device interaction. The outdoor segment, constituting 4.92\%, provided insight into the more dynamic and mobile interactions users might experience. Notably, 95.81\% of the captured images featured only one person. Regarding background objects, we discovered that walls, lights, pictures, and windows were the most common elements. The presence of terms such as "pillow" could imply individuals reclining, while words like "plant," "moon," "flower," and "cloud" might suggest outdoor settings. Overall, it appears that a significant number of images were captured indoors against plain backdrops, possibly within homes or offices. To visually represent these background objects, we have created a word cloud, which can be seen in Figure~\ref{fig:backgroundobjects}.

\begin{table}[ht!]  
\caption{{Image Characteristics: Different characteristics of the image captured, such as image angle, dominant colors, lighting conditions, photo location and number of people present. The accuracy and Cohen's kappa are presented in braces next to the categories. These results indicate substantial agreement between the annotators and alignment with VQA responses, indicating high reliability, consistency and accuracy.}}   
\begin{tabular}{@{}p{0.7\linewidth}l@{}}  
\textbf{Characteristics} & \textbf{Count} \\ \midrule  
\multicolumn{2}{l}{\cellcolor[HTML]{EFEFEF}\textit{Image Angle} (Acc=96.67; $\kappa$=0.82)} \\  
{Low angle}\begin{imageonly}{\hspace{5mm}\color{black}\rule[0.1ex]{24mm}{4pt}}\end{imageonly} & 105,949 (96.08\%)   \\
{High angle}\hspace{-0.70mm}\begin{imageonly}{\hspace{5mm}\color{black}\rule[0.1ex]{2mm}{4pt}}\end{imageonly}  & 4,323 (3.92\%) \\ \midrule  
\multicolumn{2}{l}{\cellcolor[HTML]{EFEFEF}\textit{Dominant Colors} (Acc=96.81; $\kappa$=0.75)} \\  
White \hspace{4.5mm}\begin{imageonly}{\hspace{5mm}\color{black}\rule[0.1ex]{16mm}{4pt}}\end{imageonly} & 744,14 (67.51\%) \\  
Black \hspace{5.4mm}\begin{imageonly}{\hspace{5mm}\color{black}\rule[0.1ex]{2mm}{4pt}}\end{imageonly} & 9,586 (8.70\%) \\  
Brown \hspace{4.0mm}\begin{imageonly}{\hspace{5mm}\color{black}\rule[0.1ex]{1mm}{4pt}}\end{imageonly} & 6,053 (5.49\%) \\  
Blue \hspace{6.8mm}\begin{imageonly}{\hspace{5mm}\color{black}\rule[0.1ex]{1mm}{4pt}}\end{imageonly} & 5,809 (5.27\%) \\  
Gray \hspace{6.3mm}\begin{imageonly}{\hspace{5mm}\color{black}\rule[0.1ex]{1mm}{4pt}}\end{imageonly} & 5,197 (4.72\%) \\  
Other \hspace{5.2mm}\begin{imageonly}{\hspace{5mm}\color{black}\rule[0.1ex]{2mm}{4pt}}\end{imageonly} & 9,213 (8.31\%) \\\midrule
\multicolumn{2}{l}{\cellcolor[HTML]{EFEFEF}\textit{Lighting Conditions} (Acc=89.03; $\kappa$=0.81)} \\  
Well lit \hspace{3.4mm}\begin{imageonly}{\hspace{5mm}\color{black}\rule[0.1ex]{20mm}{4pt}}\end{imageonly} & 88,843 (80.57\%)   \\
Dimly lit \hspace{1.3mm}\begin{imageonly}{\hspace{5mm}\color{black}\rule[0.1ex]{2.5mm}{4pt}}\end{imageonly} & 11,418 (10.35\%) \\
Poorly lit \hspace{0.9mm}\begin{imageonly}{\hspace{5mm}\color{black}\rule[0.1ex]{2mm}{4pt}}\end{imageonly}& 10,011 (9.08\%)\\ \midrule
\multicolumn{2}{l}{\cellcolor[HTML]{EFEFEF}\textit{Photo Location} (Acc=98.27; $\kappa$=0.71)} \\  
Indoors \hspace{2.9mm}\begin{imageonly}{\hspace{5mm}\color{black}\rule[0.1ex]{23mm}{4pt}}\end{imageonly} & 104,800 (95.08\%)   \\
Outdoors \hspace{0.8mm}\begin{imageonly}{\hspace{5mm}\color{black}\rule[0.1ex]{2.5mm}{4pt}}\end{imageonly} & 5,472 (4.92\%) \\
\multicolumn{2}{l}{\cellcolor[HTML]{EFEFEF}\textit{No. of People in the Image }(Acc=97.89; $\kappa$=0.75)} \\  
One\hspace{8.4mm}\begin{imageonly}{\hspace{5mm}\color{black}\rule[0.1ex]{23mm}{4pt}}\end{imageonly} & 105,657 (95.81\%)   \\
Two\hspace{8.3mm}\begin{imageonly}{\hspace{5mm}\color{black}\rule[0.1ex]{0.5mm}{4pt}}\end{imageonly} & 523 (0.47\%) \\
Three + \hspace{3.1mm}\begin{imageonly}{\hspace{5mm}\color{black}\rule[0.1ex]{0.1mm}{4pt}}\end{imageonly} & 8 (0.01\%) \\
None \hspace{6.1mm}\begin{imageonly}{\hspace{5mm}\color{black}\rule[0.1ex]{2mm}{4pt}}\end{imageonly} & 4084 (3.71\%) \\
\bottomrule  
\end{tabular}  
\label{tbl:communication_style}  
\end{table}  

\subsection{Predictive Analysis} \label{subsec:predictive_analysis}

\begin{table*}[ht!]
\caption{Performance: Depression detection using machine learning and deep learning methods. Standard deviation is given in braces. `LR + EN' refers to logistic regression for depression classification and elastic net for regression i.e., raw PHQ-8 score prediction. \textcolor{black}{$\pmb{R^2}$ values are presented in Appendix \ref{sec:r2}.}}
\small
\begin{tabular}{@{}llll@{}}
\toprule
\textbf{Method} &\textbf{Balanced Accuracy} $\uparrow$ &\textbf{MCC} $\uparrow$ & \textbf{MAE} $\downarrow$ \\ \midrule
Baseline & 0.52 (0.05) &0.03 (0.11) & 138.18 (3.71) \\ \cdashline{0-3}
LR + EN (MI) &0.52 (0.02) &0.04 (0.03) &135.40 (2.65) \\
Random Forest (MI) &0.54 (0.02) &0.06 (0.04) &134.45 (2.05) \\
Random Forest (3D Landmarks) &0.60 (0.04) &\textbf{0.14 (0.08)} &\textbf{130.31 (3.94)} \\ 
EffNet &\textbf{0.61 (0.02)} & 0.03 (0.00) & 137.19 (3.67) \\
\bottomrule
\end{tabular}

\label{tab:ml_and_dl}
\end{table*}

In our analysis, we leveraged both machine learning and deep learning to assess MoodCapture's ability to detect depression in natural settings. As shown in Table \ref{tab:ml_and_dl}, the EffNet model shows better performance in correctly identifying classes, as evidenced by its 0.61 balanced accuracy. However, it is interesting to note that the RF model outperforms the deep learning model in terms of overall classification (MCC of 0.14) and regression task i.e., predicting PHQ-8 scores (with an MAE of 130.31). Notably, RF achieves a lower MAE than the baseline model (130.31 vs. 138.18), indicating an approximately 6\% improvement. This makes it more robust for handling measurement errors when setting ground truth thresholds for classification.

{Furthermore, we gain several modeling insights from Table \ref{tab:ml_and_dl}. First, we observe that RF outperforms LR across all metrics, suggesting that decision trees with bagging are useful in modelling face features for depression. RF's ability to model non-linear dependencies and in-built feature selection makes it a good candidate for our problem. Second, we notice that manual feature selection, such as using 3D landmarks offer better performance than using automatic feature selection methods with MI. This finding underscores the importance of conducting an ablation study to determine the most impactful features for our analysis.}
  
{In summary, an RF trained with 3D landmarks performs well across both classification and regression tasks indicated by well-balanced scores across balanced accuracy (0.60), MCC (0.14), and MAE (130.31). Moreover, RF offers better explainability compared to deep learning methods, making it an ideal choice for post-hoc analysis (Section \ref{subsec:feature_importance}).} Our investigation into depression detection and PHQ-8 prediction using machine learning and deep learning methods provides important insights into the potential of different techniques when applied to MoodCapture data in naturalistic conditions. The results emphasize the importance of considering a range of methods, from deep learning models capable of learning complex features to traditional machine learning techniques that offer interpretability and simplicity. By carefully selecting and fine-tuning these models, we can improve the overall performance and applicability of depression detection systems in real-world scenarios.
 
\subsection{Ablation Study} \label{subsec:ablation_study}  
In this analysis, we aimed to determine if specific OpenFace feature sets are more useful for depression detection by evaluating the performance across the seven groups (Facial action units, Gaze, Eye landmarks, Pose, Rigidity Parameters, 2D and 3D landmarks). From Table \ref{tab:ablation_study}, we make several interesting observations that provide insights into the utility of individual feature sets.  
   
First, we notice that many feature sets perform better than the automatic feature selection using MI, indicating that only some specific features in the image are useful for depression detection. This finding suggests that a more focused approach to feature extraction and selection may improve overall performance. Second, we observe that facial action units are less discriminative than other features. This result may be attributed to the presence of partial face images, which are common in front-facing cameras, thus hindering the effectiveness of action units in detecting depression. Third, we find that gaze features outperform eye landmarks, suggesting that gaze direction and angle are useful. These observations highlight the importance of capturing subtle facial changes when developing depression detection systems.  

\begin{table}[ht!]
\caption{{Ablation Study: Investigating depression detection of OpenFace feature sets using a \textit{random forest}. The standard deviation is presented in braces.} \textcolor{black}{$\pmb{R^2}$ values are presented in Appendix \ref{sec:r2}.}}
\small\tabcolsep3.5pt
\begin{tabular}{@{}llll@{}}
\toprule
\textbf{Feature Set} &\textbf{Balanced Accuracy} $\uparrow$ &\textbf{MCC} $\uparrow$ & \textbf{MAE} $\downarrow$ \\ \midrule
Facial Action Units &0.54 (0.02) &0.06 (0.04) &133.34 (3.02) \\
Gaze &0.55 (0.02) &0.10 (0.05) & 132.57 (3.11) \\
Eye Landmarks &0.54 (0.02) &0.08 (0.04) &132.04 (2.90) \\
Head Pose &0.55 (0.03) &0.11 (0.05) &131.01(3.68) \\
Rigidity Parameters &0.55 (0.05) &0.06 (0.04) &133.11 (3.08) \\
2D Landmarks &0.53 (0.03) &0.05 (0.06) &132.62 (3.41) \\
3D Landmarks &\textbf{0.60 (0.04)} &\textbf{0.14 (0.08)} &\textbf{130.31 (3.94)} \\
\bottomrule
\end{tabular}
\label{tab:ablation_study}
\end{table}

{Table \ref{tab:ablation_study} also indicates that 3D Landmarks is the best performing feature set for classification and PHQ-8 prediction across all metrics (balanced accuracy=0.60; MCC=0.14; MAE=130.31). \textcolor{black}{These results suggest that an RF trained with 3D landmarks is more accurate, correlates better with the ground truth, and has lower PHQ-8 prediction errors than other methods.} 3D landmarks (see Table 2) are coordinates of specific points on the face. For example, a point in the right eye is represented as landmark number 38. This location is represented using coordinates. All landmark numbers are described in~\cite{sagonas2013300, sagonas2016300}. Intuitively, different values of 3D landmarks correspond to changes in facial expressions over time.}

In conclusion, the ablation study provides valuable insights into the utility of specific feature sets for depression detection. By understanding the strengths and limitations of individual features, researchers and practitioners can make informed decisions when designing and implementing depression detection systems, ultimately improving overall performance and applicability in real-world scenarios.

\subsection{{Machine Learning Feature Importance}} \label{subsec:feature_importance} 

{It is crucial to understand important face features that are correlated with depression. Therefore, we employ a post-hoc explainability approach, namely SHapley Additive exPlanations (SHAP) \cite{lundberg2017unified}, to investigate our best performing (Table \ref{tab:ml_and_dl}) Random Forest model. SHAP explains the model outputs using notions from game theory. It assigns each feature an importance value for a particular prediction, offering insights into how and why a model makes its decisions.}  

\begin{figure}
     \centering
     \begin{subfigure}[b]{0.49\textwidth}
         \centering
         \includegraphics[width=\textwidth]{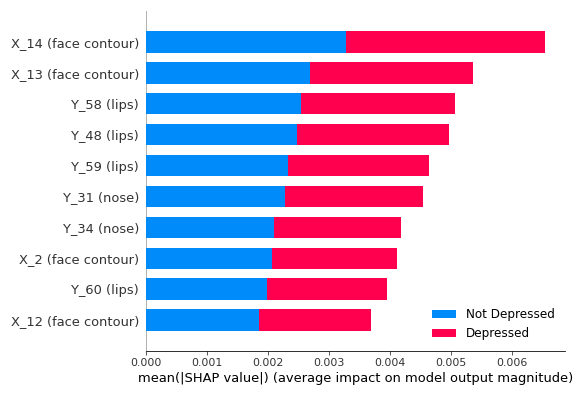}
         \caption{Important features for depression classification.}
         \label{fig:shap_classification}
     \end{subfigure}
     \hfill
     \begin{subfigure}[b]{0.49\textwidth}
         \centering
         \includegraphics[width=\textwidth]{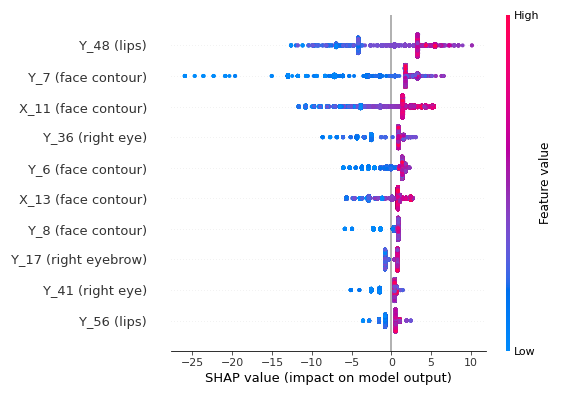}
         \caption{Important features for predicting raw depression score.}
         \label{fig:shap_regression}
     \end{subfigure}
        \caption{{SHAP plots describing the top 10 features for the classification and regression tasks. The best performing random forest trained using 3D landmark features is evaluated using SHAP. The features are x and y axis with the numbers (0-indexed) corresponding to facial landmarks \cite{sagonas2013300, sagonas2016300}}.}
        \label{fig:shap}
             \Description{Figure shows the SHAP plots for feature interpretability. The top figure shows the top 10 features for the classification whereas bottom figure  shows the top 10 features for regression task i.e. when predicting raw PHQ 8 score. Face contour, lips and nose related features are more frequently identified as important features.}
\end{figure}

The top ten important features for depression classification and regression are shown in Figure \ref{fig:shap}. Here, we observe that lips and face contour position are useful for both depression classification and score prediction. For instance, we notice that larger values of face contour near the left cheek (X\_14, X\_13, X\_11) influence the model towards predicting depression and push the raw PHQ-8 score higher. Interestingly, we find that important eye and lip features occur on the right side of the face (Y\_48, Y\_36, Y\_17, Y\_41); and higher values are associated with higher depression scores. This indicates that the ML model captures asymmetry associated with front-facing camera pictures, i.e., the right side of the face could be more visible. We discuss this further in Section \ref{sec:discussion}.

\subsection{{Investigating Bias in Machine Learning}} \label{subsec:model_bias}

\begin{figure}
     \centering
     \begin{subfigure}[b]{0.49\textwidth}
         \centering
         \includegraphics[width=\textwidth]{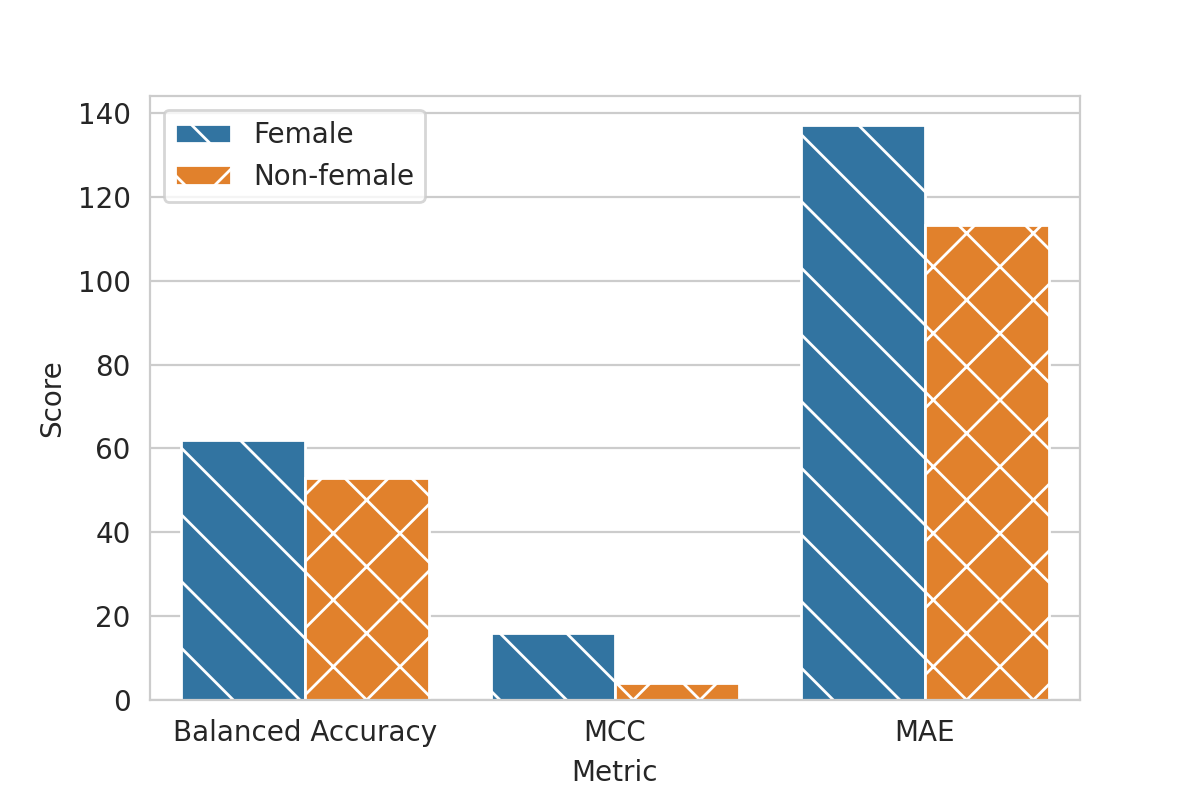}
         \caption{Gender-wise performance comparison}
         \label{fig:gender_bias}
     \end{subfigure}
     \hfill
     \begin{subfigure}[b]{0.49\textwidth}
         \centering
         \includegraphics[width=\textwidth]{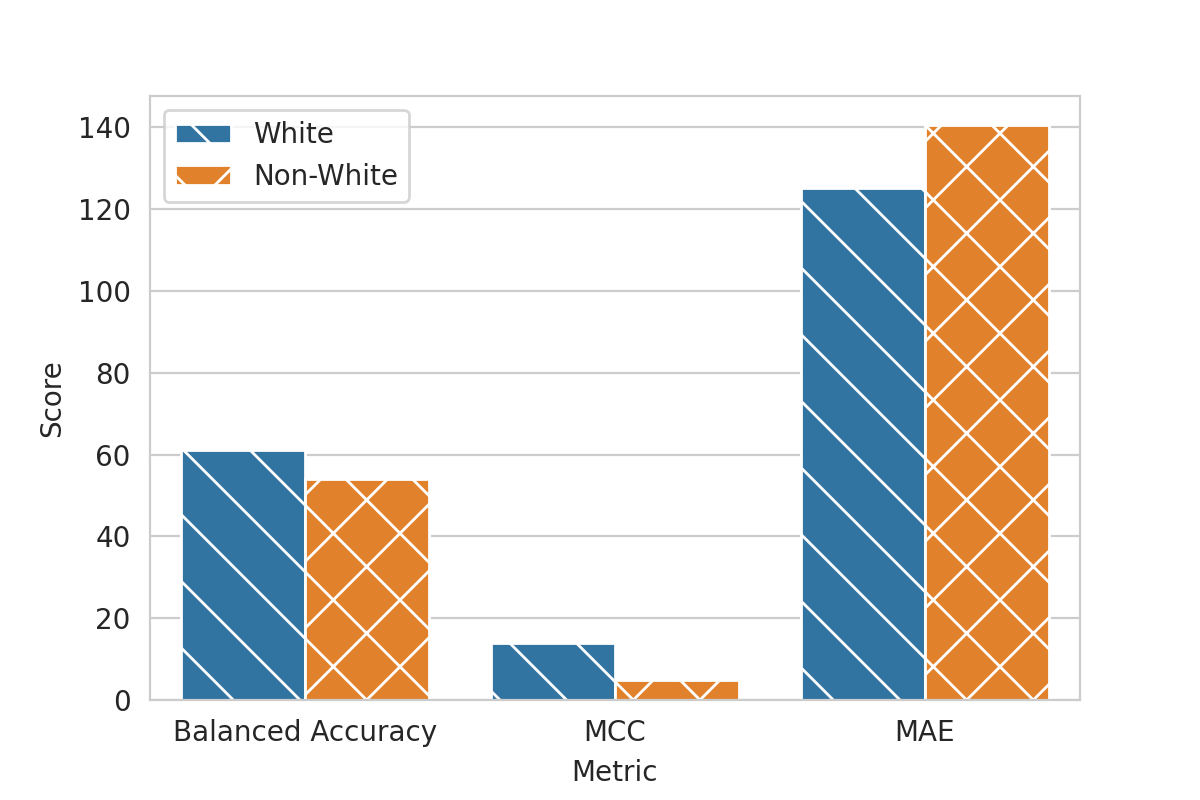}
         \caption{Race-wise performance comparison}
         \label{fig:race_bias}
     \end{subfigure}
        \caption{{Random forest performance on sub-populations divided by gender and race. Note that balanced accuracy and MCC are multiplied by 100 for better visualization.}}
        \label{fig:model_bias}
        \Description{Figure shows the result of bias analysis of the models. The performance (i.e., balanced accuracy, mcc and mae) of random forest model on sub-populations divided by their gender and race in the top figure and the bottom figure as a bar plot, respectively. Depression classification is more accurate for females, whereas regression is more precise for non-females as they have lower MAE. In terms of race, the model is more effective for white participants for both classification and regression tasks.}
\end{figure}

Our dataset predominantly consists of white females, highlighting the need to assess biases in our machine learning models related to gender and race. Thus, we categorize the test dataset into two groups for gender: females, and a combined group of males and non-binary individuals. We made the decision to combine the groups due to the notably smaller representation of non-binary individuals and males in our study. This decision aimed to address the imbalance and ensure a more meaningful analysis, acknowledging the constraints posed by the limited sample sizes of these specific demographic groups. Similarly, we classify the data into white and non-white categories for race. Again, this binary grouping strategy is designed to increase group sizes, thereby improving the statistical power of our analysis. We use our best performing RF model for these evaluations.

Figure \ref{fig:model_bias} displays the performance results of our models, revealing several notable observations. \textcolor{black}{Firstly, as indicated by MCC scores, we observe that the classifier predictions show some correlation with the ground truth at varying levels across different genders and races.} Secondly, as shown in Figure \ref{fig:gender_bias}, the results for depression classification and regression varied by gender. Specifically, we found that depression classification was more accurate for females, whereas PHQ-8 score predictions were more precise for non-females, as indicated by their lower MAE. Thirdly, the race-based performance analysis in Figure \ref{fig:race_bias} demonstrated that the model was more effective for white participants in both the classification and regression tasks. These biases likely stem from the predominance of white females in our dataset, a limitation that we discuss in Section \ref{sec:limitations}. Our analysis of these biases is intended to enhance transparency in machine learning models, providing insights for future research in this area.

\section{Ethical Considerations and User Acceptance}
\label{sec:ethical_considerations}

\begin{figure}[h]
    \centering
\includegraphics[trim={0cm 0cm 0.5cm 1cm},clip,width=1\linewidth]{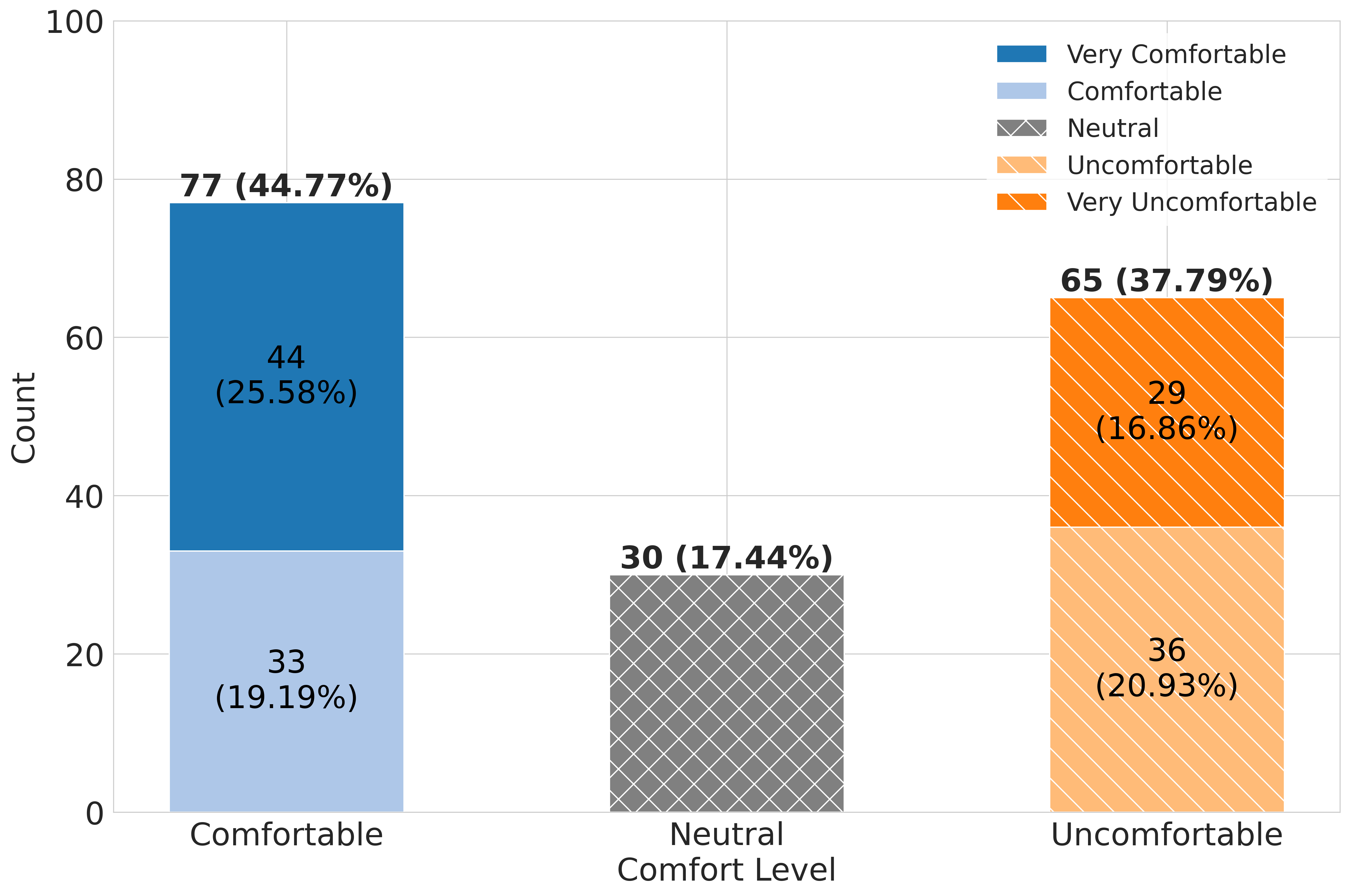}
    \caption{Comfort Level: Participant's comfort with the automated capture of their photos.}
    \label{fig:comfort}
        \Description{Figure shows stacked bar plot of participants' comfort level in having their photo captured during the study. 44.77\% of participants reported either very comfortable or comfortable, 17.44\% reported neutral and 37.79\% reported either uncomfortable or very uncomfortable.}
\end{figure}

In studies involving sensitive mental health data, it is paramount to address the ethical implications to safeguard participants' privacy, confidentiality, and well-being. Our primary goal was to prioritize the security and confidentiality of the data. We securely stored all collected data and granted access only to specific team members. We took great care in removing all personally identifiable information by implementing a thorough anonymization process. To respect privacy, any image that unintentionally captured subjects or nudity was identified during a review by two team members and subsequently deleted. We understand the sensitive nature of mental health and made sure to maintain transparency with our participants. They were informed about the study's purpose, methodology, and expected outcomes. This approach not only sought their permission but also ensured they felt comfortable and safe throughout the process. We further clarified that their compensation was unrelated to their photos.

     At the end of the study, we asked participants about their comfort levels with automated front-facing photo capture during surveys. This was optional, so we have responses from only 172 out of the 181 participants that were recruited.  Approximately 45\% of participants were comfortable, while 38\% felt it was intrusive or uneasy, and the remaining 17\% were neutral. \textcolor{black}{If participants were uncomfortable, we further ask them about specific reasons for their feelings which can be summarized into a few key themes, as shown below.} While we acknowledge these concerns, it is important to note that the study followed strict privacy and data protection guidelines.
\begin{enumerate}
    \item \textbf{Privacy and Surveillance:} Participants felt uncomfortable with the idea of being watched or monitored, as it evoked a sense of intrusion into their personal space. One participant mentioned, \textit{``I don't like being watched. I'm already paranoid when it comes to cameras.''}
    \item \textbf{Appearance and Self-Esteem:} Several participants mentioned their discomfort with having their photos taken due to concerns about their appearance. One participant stated, \textit{``I don't want people to see photos of me''}, while another said, \textit{``I am very uncomfortable with my appearance when I'm depressed.''}
    \item \textbf{Inappropriate Situations:} Participants worried about the possibility of photo bursts being taken during inconvenient or inappropriate moments. One participant shared, \textit{``If I was comfortable and at home, during some of them I may not have been completely covered.''}
    \item \textbf{Data Security:} Although participants were aware of the study's data protection measures, some still expressed concerns about the safety and storage of their images. One participant expressed, \textit{``The idea of my picture being out there ...although I know it was to be analyzed with AI.''}
    \item \textbf{Lack of Control:} Participants felt uneasy about not being able to review, approve, or delete the photos taken during the bursts, as well as not knowing when the camera was active. A participant shared, \textit{``Having pictures taken and not knowing what they looked like or if they were embarrassing is an uncomfortable thing to think about.''}
\end{enumerate}

In summary, participants' concerns mainly revolved around privacy, self-esteem, potential inappropriate situations, data security, and control over the images. It is essential to consider these concerns when designing and implementing studies involving photo bursts or similar data collection methods to ensure participants' comfort and trust in the research process. Acknowledging the sensitive nature of our research, we offered participants the option to delete their photos at the end of the study if they felt uncomfortable. Interestingly, no participants chose this option, highlighting the trust they placed in our research process and commitment to ethical conduct. We remain keenly aware of the potential for technology misuse, especially in unauthorized surveillance or data mining scenarios. We have taken measures to minimize such risks, emphasizing that our technological developments are primarily intended as health aids, not tools for unwarranted monitoring. Further, to address participants' concerns regarding privacy and data security, one possible solution could be leveraging the capabilities of AI chips on smartphones. By conducting all image classification and processing on the device itself, no images would need to be transmitted or stored externally. This approach could significantly alleviate users' concerns about their images being stored or accessed by unauthorized parties. As AI technology continues to advance, incorporating on-device processing capabilities into our research methodology may not only increase user trust and comfort but also pave the way for a new generation of privacy-focused health aids. In line with our commitment to ethical conduct, we will continue to explore and implement such technological advancements to ensure the protection of participants' data and privacy in our research. 

\section{Discussion}
\label{sec:discussion}
\textcolor{black}{In this section, we provide a summary of our findings and engage in a thorough discussion, exploring the implications and uncovering the potential opportunities highlighted by our results.}
\subsection{Summary of results}
\label{sec:summary}

Our study investigated the potential of using in-the-wild smartphone images and deep learning models for detecting depression and predicting PHQ-8 scores, aiming to contribute to the development of user-centered and unobtrusive mental health assessment tools. The results of our analysis provided valuable insights into the characteristics of in-the-wild images, the performance of machine learning and deep learning models, and user acceptance of such approaches. The image characteristics analysis revealed that most images were captured from a low angle, indoors, and under well-lit conditions. These findings highlighted the participants' natural behavior with their smartphones, emphasizing the importance of considering real-world HCI dynamics in designing mental health assessment tools.

{Our predictive analysis demonstrates that a random forest model trained by manually selecting 3D landmark features obtains the best overall classification (balanced accuracy of 0.60, MCC of 0.14) and regression performance (MAE of 130.31). Interestingly, the EffNet deep learning model barely beat this score for classification task by 0.01. It correctly identified depressed and non-depressed participants with a balanced accuracy of 0.61. Given additional high quality data, the deep learning models could improve over existing methods. To summarize, these scores are promising.} They are even more noteworthy considering that the facial images were captured using a diverse range of smartphone devices – 87 different models from 9 distinct brands. As the camera quality of these devices varies significantly, it is important to note that the results may be influenced by factors such as image clarity and auto-focus capabilities. Despite these potential limitations, our findings support the ecological validity of the study and emphasize the potential of machine learning and deep learning methods in analyzing depression from facial images, even when captured in less-than-ideal conditions. 

{During post-hoc analysis, we gained several interesting insights. Firstly, our ablation study indicates that smaller domain-specific feature sets perform better in both our tasks. Specifically, we notice that 3D landmarks, gaze, and pose offer good performance across all metrics.} By focusing on these features, researchers can potentially improve the overall performance of mental health assessment tools. {Secondly, our explainability analysis revealed that larger values on the right side of the face have an impact on both depression detection and PHQ-8 score prediction. \textcolor{black}{This finding suggests that people hold phones in a way that emphasizes the asymmetry of front-facing face images.} Thirdly, our investigation into biases within machine learning models offers crucial insights for future research, particularly in terms of improving generalization and guiding data collection strategies.} In terms of user acceptance, we found diverse responses regarding participants' comfort levels with automated front-facing photo capture. While some participants were comfortable with the process, others felt uneasy due to concerns related to privacy, self-esteem, inappropriate situations, data security, and control over the images. These concerns highlight the need for careful consideration of ethical implications in designing and implementing studies involving photo bursts or similar data collection methods. 

In conclusion, our research highlights the potential of using in-the-wild smartphone images, machine learning and deep learning models for depression analysis, offering a more objective, unobtrusive, and continuous approach to mental health assessment. By carefully considering the insights gained from our analysis and addressing the ethical implications, researchers and practitioners can work towards developing user-centered, effective, and ethically sound tools for mental health assessment and intervention.

\subsection{Implications}
The findings from our study hold significant implications for various stakeholders, including researchers, practitioners, and policymakers in the fields of mental health, digital health, human-computer interaction (HCI), and public health.

Our research highlights the potential of utilizing smartphone images and machine learning models as a supplementary method for mental health assessment. This innovative approach encourages the exploration of alternative ways to assess mental health that can complement traditional tools such as self-report questionnaires and clinical interviews. While our data was collected from participants who had major depressive disorder, the results pave the way for future research to investigate the broader applicability of these methods, potentially leading to a better understanding of depression and improved mental health support over time. Consequently, promoting timely access to appropriate interventions and support systems. 

From an HCI perspective, our study underscores the importance of considering user acceptance when developing mental health assessment tools that utilize smartphone images and machine learning. {Recently, there has been a growing interest among researchers to integrate user acceptance into the training phase of machine learning models, as proposed in studies like \cite{lobner2023user, bouneffouf2013applying}. In a related observation, our feature importance analysis indicated that the right side of the face is more useful in depression detection. This phenomenon could be linked to the dominance of right-handed individuals, often resulting in partial face images that capture more of the right side. Various studies support the idea that handedness influences user interaction with smartphones and user experience (UX) \cite{acscci2014left, nelavelli2018adaptive, golles2017usability, kurniawan2020design}. Therefore, future research in HCI could benefit from focusing on developing tools that facilitate the capture of the entire face more effectively. For instance, the work by \citet{nelavelli2018adaptive} explores adaptive app design tailored to the user's handedness, which could be a promising direction for enhancing face image capture in smartphone applications.} In summary, understanding users' concerns and preferences is crucial for creating tools that are more likely to be adopted and used by those in need of support. This focus on user acceptance can inspire the HCI community to design mental health assessment tools that balance effectiveness, privacy, and user engagement, leading to the development of more accessible and inclusive digital mental health solutions.

In the broader context of public health, the study's findings emphasize the importance of leveraging technology and innovative methods to address mental health challenges. As mental health disorders continue to impact individuals and communities worldwide, adopting novel approaches like the one presented in our study can contribute to more effective prevention strategies, early intervention, and resource allocation. This could ultimately lead to better mental health outcomes and overall well-being for individuals across various demographic and cultural contexts. In summary, the implications of our study extend well beyond the immediate findings, offering valuable insights for a range of stakeholders working at the intersection of mental health, digital health, and human-computer interaction. By considering user acceptance, exploring the potential of smartphone images for mental health assessment, and recognizing the broader public health context, our study contributes to the development of more effective, user-friendly, and contextually appropriate mental health assessment tools with the potential to improve the lives of individuals affected by depression.

\section{Limitations}
\label{sec:limitations}
Our study while providing valuable insights into the use of in-the-wild smartphone images and deep learning models for depression detection, has some limitations that should be acknowledged. First, our study's dataset may be limited in size and diversity, as it consists of a relatively small number of participants. {Additionally, it is important to remember that our dataset is primarily composed of white females. Although our models currently show better performance for females in classification tasks and for non-females in regression tasks, expanding our dataset to include more diverse samples is necessary. By incorporating additional data that represents a broader spectrum of the population, we can ensure a more comprehensive representation. This expansion will not only enhance the robustness of our findings but also significantly improve the generalizability of our results across different demographic groups.} Furthermore, the study relies on self-reported data, such as depression scores, which may be subject to biases, including social desirability and recall bias. Future research could be significantly enhanced by including more objective measures of mental health, such as clinical evaluations or physiological indicators. {In our study, we adjusted each item's score on the PHQ-8 from its original 0-3 range to a broader 0-100 scale. As mentioned earlier, the practice of re-scaling psychometric scales is not uncommon and has been applied to the PHQ in various past studies~\cite{Majethia2022, Nguyen2021, Gumus2023, Lum2016}. However, one limitation of adapting the PHQ-8 to a 0-100 scale is the potential for inconsistencies when correlating these scores with established levels of depression severity. To mitigate this, we proportionally scaled the original scores to derive our depression categorization, striving to preserve the original scoring system's integrity. Additionally, our prediction models consider both the raw PHQ scores and the adjusted class scales, an approach that aims to balance detailed granularity with traditional scoring validity.}  
It is also important to highlight that all participants in our study had received clinical diagnoses for MDD. However, we relied on self-reported data for tracking daily depression levels, which facilitated more consistent monitoring. Our study also focused exclusively on a clinically depressed cohort. Including healthy individuals in the dataset would have been beneficial for developing a more comprehensive and accurate prediction model. A randomized controlled trial (RCT) with healthy controls or incorporating a diverse cohort of individuals not experiencing depression could provide valuable insights into the differences between depressed and non-depressed individuals and improve the model's ability to distinguish between them. Future research should consider expanding the dataset to include both depressed and healthy individuals, which can contribute to the development of more effective and precise mental health assessment tools.

Another limitation is that the study primarily focuses on the analysis of in-the-wild smartphone images and their relationship with depression. However, there may be other factors, such as social interactions, physical activity, and environmental context, that could provide additional insights into depression detection. Integrating these factors into future research may help to develop more holistic and accurate prediction models. Deep learning models, while powerful and effective, can often be considered as "black-box" models with limited interpretability. This may make it difficult to understand the specific features or patterns that the model has identified as being related to depression. Future research could explore the use of more interpretable models or techniques to provide insights into the underlying mechanisms linking visual cues and depression. Lastly, the use of in-the-wild smartphone images for mental health assessment raises ethical and privacy concerns, which need to be carefully considered when designing and implementing such tools. Ensuring user consent, data security, and transparency in the use of personal data is crucial for maintaining trust and fostering the adoption of these tools. Addressing these limitations in future research can help to further advance our understanding of the relationship between smartphone images, deep learning models, and depression detection, contributing to the development of more effective, user-centered, and ethically sound mental health assessment tools.
\section{Conclusion and Future Work}
\label{sec:conclusion}
Through this study, we have demonstrated the potential of using in-the-wild smartphone images and machine learning to detect depression, offering valuable insights for mental health assessment, HCI and digital health. With this, we aim to pave the way for more effective and user-centered mental health assessment tools. Addressing the limitations of our study and building upon its findings, future research can contribute to the development of more robust, accurate, and ethically sound mental health assessment tools that have the potential to improve the lives of individuals affected by depression. 

When we embarked on designing our MoodCapture study to investigate whether high-resolution face capture from phones could assess mood, we were acutely aware of the ethical issues surrounding our research and the potential privacy concerns of a population that included individuals diagnosed with depression. As discussed in the section on Ethical Considerations and User Acceptance, our study was meticulously designed to safeguard user privacy throughout, and we sought their evaluations of the MoodCapture app post-study. This invaluable feedback forms the foundation for future work in image-based mood detection which we believe is a promising technology. One direction we plan to pursue as our next step involves utilizing on-phone AI chips that are now available on top-end smartphones to run deep learning models directly on the device, ensuring that images never leave the phone. Additionally, we intend to explore the combination of this on-device prediction approach with federated deep learning, where models are trained without sharing raw data across a network in a central entity such as a server or cloud. This approach could effectively address security concerns associated with centralized data collection and the privacy issues our participants raised during the acceptance study. {Finally, we recognize that the performance of the models we considered for face-based depression detection, particularly deep learning models, would benefit significantly from a larger face dataset. In the MoodCapture study, we collected over 125,000 images from 177 individuals living with depression over a period of 90 days, representing a well-sized dataset to demonstrate the potential of this idea. If future face-based depression studies have access to larger pools of naturalistic images (e.g., VGGFace2, which contains over 3 million face images) collected in the wild, we anticipate that the accuracy and capabilities of the models would see significant improvement.}

\begin{acks}
\label{sec:ack}
\textcolor{black}{We sincerely thank the participants who kindly consented to share their photos for this study. The CHI reviewers helped lift this paper considerably. Their insight, detailed comments, and suggestions were invaluable. We thank them for their diligence and for caring about the subject matter and our paper. The research discussed in this paper was supported by the National Institute of Mental Health (NIMH) under award number R01MH123482-01. We acknowledge that the content of this manuscript is solely our responsibility and does not necessarily reflect the views of the NIMH. We also clarify that the funding body had no involvement in the study's design, data collection, analysis, interpretation, or manuscript preparation.}
\end{acks}

\bibliographystyle{ACM-Reference-Format}
\bibliography{paper_bib}


\begin{thebibliography}{79}


\ifx \showCODEN    \undefined \def \showCODEN     #1{\unskip}     \fi
\ifx \showDOI      \undefined \def \showDOI       #1{#1}\fi
\ifx \showISBNx    \undefined \def \showISBNx     #1{\unskip}     \fi
\ifx \showISBNxiii \undefined \def \showISBNxiii  #1{\unskip}     \fi
\ifx \showISSN     \undefined \def \showISSN      #1{\unskip}     \fi
\ifx \showLCCN     \undefined \def \showLCCN      #1{\unskip}     \fi
\ifx \shownote     \undefined \def \shownote      #1{#1}          \fi
\ifx \showarticletitle \undefined \def \showarticletitle #1{#1}   \fi
\ifx \showURL      \undefined \def \showURL       {\relax}        \fi
\providecommand\bibfield[2]{#2}
\providecommand\bibinfo[2]{#2}
\providecommand\natexlab[1]{#1}
\providecommand\showeprint[2][]{arXiv:#2}

\bibitem[Amaltinga and Mbinta(2020)]%
        {Amaltinga2020}
\bibfield{author}{\bibinfo{person}{Awuni Prosper~Mandela Amaltinga} {and} \bibinfo{person}{James~Fenibe Mbinta}.} \bibinfo{year}{2020}\natexlab{}.
\newblock \showarticletitle{Factors associated with depression among young people globally: a narrative review}.
\newblock \bibinfo{journal}{\emph{International Journal Of Community Medicine And Public Health}} \bibinfo{volume}{7}, \bibinfo{number}{9} (\bibinfo{date}{Aug.} \bibinfo{year}{2020}), \bibinfo{pages}{3711}.
\newblock
\urldef\tempurl%
\url{https://doi.org/10.18203/2394-6040.ijcmph20203949}
\showDOI{\tempurl}


\bibitem[A{\c{s}}{\c{c}}{\i} and R{\i}zvano{\u{g}}lu(2014)]%
        {acscci2014left}
\bibfield{author}{\bibinfo{person}{Sinan A{\c{s}}{\c{c}}{\i}} {and} \bibinfo{person}{Kerem R{\i}zvano{\u{g}}lu}.} \bibinfo{year}{2014}\natexlab{}.
\newblock \showarticletitle{Left vs. right-handed UX: A comparative user study on a mobile application with left and right-handed users}. In \bibinfo{booktitle}{\emph{Design, User Experience, and Usability. User Experience Design for Diverse Interaction Platforms and Environments: Third International Conference, DUXU 2014, Held as Part of HCI International 2014, Heraklion, Crete, Greece, June 22-27, 2014, Proceedings, Part II 3}}. Springer, \bibinfo{pages}{173--183}.
\newblock


\bibitem[B{\^a}ce et~al\mbox{.}(2020)]%
        {bace2020quantification}
\bibfield{author}{\bibinfo{person}{Mihai B{\^a}ce}, \bibinfo{person}{Sander Staal}, {and} \bibinfo{person}{Andreas Bulling}.} \bibinfo{year}{2020}\natexlab{}.
\newblock \showarticletitle{Quantification of users' visual attention during everyday mobile device interactions}. In \bibinfo{booktitle}{\emph{Proceedings of the 2020 CHI Conference on Human Factors in Computing Systems}}. \bibinfo{pages}{1--14}.
\newblock


\bibitem[Baltru{\v{s}}aitis et~al\mbox{.}(2015)]%
        {baltruvsaitis2015cross}
\bibfield{author}{\bibinfo{person}{Tadas Baltru{\v{s}}aitis}, \bibinfo{person}{Marwa Mahmoud}, {and} \bibinfo{person}{Peter Robinson}.} \bibinfo{year}{2015}\natexlab{}.
\newblock \showarticletitle{Cross-dataset learning and person-specific normalisation for automatic action unit detection}. In \bibinfo{booktitle}{\emph{2015 11th IEEE International Conference and Workshops on Automatic Face and Gesture Recognition (FG)}}, Vol.~\bibinfo{volume}{6}. IEEE, \bibinfo{pages}{1--6}.
\newblock


\bibitem[Baltrusaitis et~al\mbox{.}(2013)]%
        {baltrusaitis2013constrained}
\bibfield{author}{\bibinfo{person}{Tadas Baltrusaitis}, \bibinfo{person}{Peter Robinson}, {and} \bibinfo{person}{Louis-Philippe Morency}.} \bibinfo{year}{2013}\natexlab{}.
\newblock \showarticletitle{Constrained local neural fields for robust facial landmark detection in the wild}. In \bibinfo{booktitle}{\emph{Proceedings of the IEEE international conference on computer vision workshops}}. \bibinfo{pages}{354--361}.
\newblock


\bibitem[Baltrusaitis et~al\mbox{.}(2018)]%
        {baltrusaitis2018openface}
\bibfield{author}{\bibinfo{person}{Tadas Baltrusaitis}, \bibinfo{person}{Amir Zadeh}, \bibinfo{person}{Yao~Chong Lim}, {and} \bibinfo{person}{Louis-Philippe Morency}.} \bibinfo{year}{2018}\natexlab{}.
\newblock \showarticletitle{Openface 2.0: Facial behavior analysis toolkit}. In \bibinfo{booktitle}{\emph{2018 13th IEEE international conference on automatic face \& gesture recognition (FG 2018)}}. IEEE, \bibinfo{pages}{59--66}.
\newblock


\bibitem[Beck et~al\mbox{.}(1987)]%
        {beck1987beck}
\bibfield{author}{\bibinfo{person}{Aaron~T Beck}, \bibinfo{person}{Robert~A Steer}, \bibinfo{person}{Gregory~K Brown}, {et~al\mbox{.}}} \bibinfo{year}{1987}\natexlab{}.
\newblock \bibinfo{booktitle}{\emph{Beck depression inventory}}.
\newblock \bibinfo{publisher}{Harcourt Brace Jovanovich New York:}.
\newblock


\bibitem[Belouali et~al\mbox{.}(2021)]%
        {belouali2021acoustic}
\bibfield{author}{\bibinfo{person}{Anas Belouali}, \bibinfo{person}{Samir Gupta}, \bibinfo{person}{Vaibhav Sourirajan}, \bibinfo{person}{Jiawei Yu}, \bibinfo{person}{Nathaniel Allen}, \bibinfo{person}{Adil Alaoui}, \bibinfo{person}{Mary~Ann Dutton}, {and} \bibinfo{person}{Matthew~J Reinhard}.} \bibinfo{year}{2021}\natexlab{}.
\newblock \showarticletitle{Acoustic and language analysis of speech for suicidal ideation among US veterans}.
\newblock \bibinfo{journal}{\emph{BioData mining}} \bibinfo{volume}{14}, \bibinfo{number}{1} (\bibinfo{year}{2021}), \bibinfo{pages}{1--17}.
\newblock


\bibitem[Bertolote et~al\mbox{.}(2003)]%
        {Bertolote2003}
\bibfield{author}{\bibinfo{person}{Jos{\'{e}}~Manoel Bertolote}, \bibinfo{person}{Alexandra Fleischmann}, \bibinfo{person}{Diego~De Leo}, {and} \bibinfo{person}{Danuta Wasserman}.} \bibinfo{year}{2003}\natexlab{}.
\newblock \showarticletitle{Suicide and mental disorders: do we know enough?}
\newblock \bibinfo{journal}{\emph{British Journal of Psychiatry}} \bibinfo{volume}{183}, \bibinfo{number}{5} (\bibinfo{date}{Nov.} \bibinfo{year}{2003}), \bibinfo{pages}{382--383}.
\newblock
\urldef\tempurl%
\url{https://doi.org/10.1192/bjp.183.5.382}
\showDOI{\tempurl}


\bibitem[Bouneffouf(2013)]%
        {bouneffouf2013applying}
\bibfield{author}{\bibinfo{person}{Djallel Bouneffouf}.} \bibinfo{year}{2013}\natexlab{}.
\newblock \showarticletitle{Applying machine learning techniques to improve user acceptance on ubiquitous environement}.
\newblock \bibinfo{journal}{\emph{arXiv preprint arXiv:1301.4351}} (\bibinfo{year}{2013}).
\newblock


\bibitem[Breiman(2001)]%
        {breiman2001random}
\bibfield{author}{\bibinfo{person}{Leo Breiman}.} \bibinfo{year}{2001}\natexlab{}.
\newblock \showarticletitle{Random forests}.
\newblock \bibinfo{journal}{\emph{Machine learning}}  \bibinfo{volume}{45} (\bibinfo{year}{2001}), \bibinfo{pages}{5--32}.
\newblock


\bibitem[Chancellor and De~Choudhury(2020)]%
        {chancellor2020methods}
\bibfield{author}{\bibinfo{person}{Stevie Chancellor} {and} \bibinfo{person}{Munmun De~Choudhury}.} \bibinfo{year}{2020}\natexlab{}.
\newblock \showarticletitle{Methods in predictive techniques for mental health status on social media: a critical review}.
\newblock \bibinfo{journal}{\emph{NPJ digital medicine}} \bibinfo{volume}{3}, \bibinfo{number}{1} (\bibinfo{year}{2020}), \bibinfo{pages}{43}.
\newblock


\bibitem[Chicco and Jurman(2020)]%
        {chicco2020advantages}
\bibfield{author}{\bibinfo{person}{Davide Chicco} {and} \bibinfo{person}{Giuseppe Jurman}.} \bibinfo{year}{2020}\natexlab{}.
\newblock \showarticletitle{The advantages of the Matthews correlation coefficient (MCC) over F1 score and accuracy in binary classification evaluation}.
\newblock \bibinfo{journal}{\emph{BMC genomics}} \bibinfo{volume}{21}, \bibinfo{number}{1} (\bibinfo{year}{2020}), \bibinfo{pages}{1--13}.
\newblock


\bibitem[Chikersal et~al\mbox{.}(2021)]%
        {chikersal2021detecting}
\bibfield{author}{\bibinfo{person}{Prerna Chikersal}, \bibinfo{person}{Afsaneh Doryab}, \bibinfo{person}{Michael Tumminia}, \bibinfo{person}{Daniella~K Villalba}, \bibinfo{person}{Janine~M Dutcher}, \bibinfo{person}{Xinwen Liu}, \bibinfo{person}{Sheldon Cohen}, \bibinfo{person}{Kasey~G Creswell}, \bibinfo{person}{Jennifer Mankoff}, \bibinfo{person}{J~David Creswell}, {et~al\mbox{.}}} \bibinfo{year}{2021}\natexlab{}.
\newblock \showarticletitle{Detecting depression and predicting its onset using longitudinal symptoms captured by passive sensing: a machine learning approach with robust feature selection}.
\newblock \bibinfo{journal}{\emph{ACM Transactions on Computer-Human Interaction (TOCHI)}} \bibinfo{volume}{28}, \bibinfo{number}{1} (\bibinfo{year}{2021}), \bibinfo{pages}{1--41}.
\newblock


\bibitem[Cramer et~al\mbox{.}(2016)]%
        {cramer2016major}
\bibfield{author}{\bibinfo{person}{Ang{\'e}lique~OJ Cramer}, \bibinfo{person}{Claudia~D Van~Borkulo}, \bibinfo{person}{Erik~J Giltay}, \bibinfo{person}{Han~LJ Van Der~Maas}, \bibinfo{person}{Kenneth~S Kendler}, \bibinfo{person}{Marten Scheffer}, {and} \bibinfo{person}{Denny Borsboom}.} \bibinfo{year}{2016}\natexlab{}.
\newblock \showarticletitle{Major depression as a complex dynamic system}.
\newblock \bibinfo{journal}{\emph{PloS one}} \bibinfo{volume}{11}, \bibinfo{number}{12} (\bibinfo{year}{2016}), \bibinfo{pages}{e0167490}.
\newblock


\bibitem[Darvariu et~al\mbox{.}(2020)]%
        {darvariu2020quantifying}
\bibfield{author}{\bibinfo{person}{Victor-Alexandru Darvariu}, \bibinfo{person}{Laura Convertino}, \bibinfo{person}{Abhinav Mehrotra}, {and} \bibinfo{person}{Mirco Musolesi}.} \bibinfo{year}{2020}\natexlab{}.
\newblock \showarticletitle{Quantifying the relationships between everyday objects and emotional states through deep learning based image analysis using smartphones}.
\newblock \bibinfo{journal}{\emph{Proceedings of the ACM on Interactive, Mobile, Wearable and Ubiquitous Technologies}} \bibinfo{volume}{4}, \bibinfo{number}{1} (\bibinfo{year}{2020}), \bibinfo{pages}{1--21}.
\newblock


\bibitem[De~Choudhury et~al\mbox{.}(2013)]%
        {de2013predicting}
\bibfield{author}{\bibinfo{person}{Munmun De~Choudhury}, \bibinfo{person}{Michael Gamon}, \bibinfo{person}{Scott Counts}, {and} \bibinfo{person}{Eric Horvitz}.} \bibinfo{year}{2013}\natexlab{}.
\newblock \showarticletitle{Predicting depression via social media}. In \bibinfo{booktitle}{\emph{Proceedings of the international AAAI conference on web and social media}}, Vol.~\bibinfo{volume}{7}. \bibinfo{pages}{128--137}.
\newblock


\bibitem[de~la Torre et~al\mbox{.}(2020)]%
        {AriasdelaTorre2020}
\bibfield{author}{\bibinfo{person}{Jorge~Arias de~la Torre}, \bibinfo{person}{Gemma Vilagut}, \bibinfo{person}{Antoni Serrano-Blanco}, \bibinfo{person}{Vicente Mart{\'{\i}}n}, \bibinfo{person}{Antonio~Jos{\'{e}} Molina}, \bibinfo{person}{Jose~M Valderas}, {and} \bibinfo{person}{Jordi Alonso}.} \bibinfo{year}{2020}\natexlab{}.
\newblock \showarticletitle{Accuracy of Self-Reported Items for the Screening of Depression in the General Population}.
\newblock \bibinfo{journal}{\emph{International Journal of Environmental Research and Public Health}} \bibinfo{volume}{17}, \bibinfo{number}{21} (\bibinfo{date}{Oct.} \bibinfo{year}{2020}), \bibinfo{pages}{7955}.
\newblock
\urldef\tempurl%
\url{https://doi.org/10.3390/ijerph17217955}
\showDOI{\tempurl}


\bibitem[Deady et~al\mbox{.}(2021)]%
        {Deady2021}
\bibfield{author}{\bibinfo{person}{M Deady}, \bibinfo{person}{D~A~J Collins}, \bibinfo{person}{D~A Johnston}, \bibinfo{person}{N Glozier}, \bibinfo{person}{R~A Calvo}, \bibinfo{person}{H Christensen}, {and} \bibinfo{person}{S~B Harvey}.} \bibinfo{year}{2021}\natexlab{}.
\newblock \showarticletitle{The impact of depression, anxiety and comorbidity on occupational outcomes}.
\newblock \bibinfo{journal}{\emph{Occupational Medicine}} \bibinfo{volume}{72}, \bibinfo{number}{1} (\bibinfo{date}{Oct.} \bibinfo{year}{2021}), \bibinfo{pages}{17--24}.
\newblock
\urldef\tempurl%
\url{https://doi.org/10.1093/occmed/kqab142}
\showDOI{\tempurl}


\bibitem[Ebrahimi et~al\mbox{.}(2021)]%
        {ebrahimi2021within}
\bibfield{author}{\bibinfo{person}{Omid~V Ebrahimi}, \bibinfo{person}{Julian Burger}, \bibinfo{person}{Asle Hoffart}, {and} \bibinfo{person}{Sverre~Urnes Johnson}.} \bibinfo{year}{2021}\natexlab{}.
\newblock \showarticletitle{Within-and across-day patterns of interplay between depressive symptoms and related psychopathological processes: a dynamic network approach during the COVID-19 pandemic}.
\newblock \bibinfo{journal}{\emph{BMC medicine}} \bibinfo{volume}{19}, \bibinfo{number}{1} (\bibinfo{year}{2021}), \bibinfo{pages}{1--17}.
\newblock


\bibitem[Fergusson and Woodward(2002)]%
        {Fergusson2002-nc}
\bibfield{author}{\bibinfo{person}{David~M Fergusson} {and} \bibinfo{person}{Lianne~J Woodward}.} \bibinfo{year}{2002}\natexlab{}.
\newblock \showarticletitle{Mental health, educational, and social role outcomes of adolescents with depression}.
\newblock \bibinfo{journal}{\emph{Arch. Gen. Psychiatry}} \bibinfo{volume}{59}, \bibinfo{number}{3} (\bibinfo{date}{March} \bibinfo{year}{2002}), \bibinfo{pages}{225--231}.
\newblock


\bibitem[Francese and Attanasio(2022)]%
        {Francese2022}
\bibfield{author}{\bibinfo{person}{Rita Francese} {and} \bibinfo{person}{Pasquale Attanasio}.} \bibinfo{year}{2022}\natexlab{}.
\newblock \showarticletitle{Emotion detection for supporting depression screening}.
\newblock \bibinfo{journal}{\emph{Multimedia Tools and Applications}} \bibinfo{volume}{82}, \bibinfo{number}{9} (\bibinfo{date}{Dec.} \bibinfo{year}{2022}), \bibinfo{pages}{12771–12795}.
\newblock
\showISSN{1573-7721}
\urldef\tempurl%
\url{https://doi.org/10.1007/s11042-022-14290-0}
\showDOI{\tempurl}


\bibitem[Frerichs et~al\mbox{.}(1982)]%
        {Frerichs1982}
\bibfield{author}{\bibinfo{person}{Ralph~R. Frerichs}, \bibinfo{person}{Carol~S. Aneshensel}, \bibinfo{person}{Patricia~A. Yokopenic}, {and} \bibinfo{person}{Virginia~A. Clark}.} \bibinfo{year}{1982}\natexlab{}.
\newblock \showarticletitle{Physical health and depression: An epidemiologic survey}.
\newblock \bibinfo{journal}{\emph{Preventive Medicine}} \bibinfo{volume}{11}, \bibinfo{number}{6} (\bibinfo{date}{Nov.} \bibinfo{year}{1982}), \bibinfo{pages}{639--646}.
\newblock
\urldef\tempurl%
\url{https://doi.org/10.1016/0091-7435(82)90026-3}
\showDOI{\tempurl}


\bibitem[Fried et~al\mbox{.}(2022)]%
        {fried2022revisiting}
\bibfield{author}{\bibinfo{person}{Eiko~I Fried}, \bibinfo{person}{Jessica~K Flake}, {and} \bibinfo{person}{Donald~J Robinaugh}.} \bibinfo{year}{2022}\natexlab{}.
\newblock \showarticletitle{Revisiting the theoretical and methodological foundations of depression measurement}.
\newblock \bibinfo{journal}{\emph{Nature Reviews Psychology}} \bibinfo{volume}{1}, \bibinfo{number}{6} (\bibinfo{year}{2022}), \bibinfo{pages}{358--368}.
\newblock


\bibitem[Fried and Nesse(2015)]%
        {fried2015depression}
\bibfield{author}{\bibinfo{person}{Eiko~I Fried} {and} \bibinfo{person}{Randolph~M Nesse}.} \bibinfo{year}{2015}\natexlab{}.
\newblock \showarticletitle{Depression is not a consistent syndrome: An investigation of unique symptom patterns in the STAR* D study}.
\newblock \bibinfo{journal}{\emph{Journal of affective disorders}}  \bibinfo{volume}{172} (\bibinfo{year}{2015}), \bibinfo{pages}{96--102}.
\newblock


\bibitem[Garimella et~al\mbox{.}(2016)]%
        {garimella2016social}
\bibfield{author}{\bibinfo{person}{Venkata Rama~Kiran Garimella}, \bibinfo{person}{Abdulrahman Alfayad}, {and} \bibinfo{person}{Ingmar Weber}.} \bibinfo{year}{2016}\natexlab{}.
\newblock \showarticletitle{Social media image analysis for public health}. In \bibinfo{booktitle}{\emph{Proceedings of the 2016 CHI Conference on Human Factors in Computing Systems}}. \bibinfo{pages}{5543--5547}.
\newblock


\bibitem[Golles(2017)]%
        {golles2017usability}
\bibfield{author}{\bibinfo{person}{Filip~Norman Golles}.} \bibinfo{year}{2017}\natexlab{}.
\newblock \showarticletitle{Usability of mobile interfaces with regards to left-handed use}.
\newblock \bibinfo{journal}{\emph{USCCS 2017}} (\bibinfo{year}{2017}), \bibinfo{pages}{65}.
\newblock


\bibitem[Gong and Poellabauer(2017)]%
        {gong2017topic}
\bibfield{author}{\bibinfo{person}{Yuan Gong} {and} \bibinfo{person}{Christian Poellabauer}.} \bibinfo{year}{2017}\natexlab{}.
\newblock \showarticletitle{Topic modeling based multi-modal depression detection}. In \bibinfo{booktitle}{\emph{Proceedings of the 7th annual workshop on Audio/Visual emotion challenge}}. \bibinfo{pages}{69--76}.
\newblock


\bibitem[Gumus et~al\mbox{.}(2023)]%
        {Gumus2023}
\bibfield{author}{\bibinfo{person}{Melisa Gumus}, \bibinfo{person}{Danielle~D DeSouza}, \bibinfo{person}{Mengdan Xu}, \bibinfo{person}{Celia Fidalgo}, \bibinfo{person}{William Simpson}, {and} \bibinfo{person}{Jessica Robin}.} \bibinfo{year}{2023}\natexlab{}.
\newblock \showarticletitle{Evaluating the utility of daily speech assessments for monitoring depression symptoms}.
\newblock \bibinfo{journal}{\emph{DIGITAL HEALTH}}  \bibinfo{volume}{9} (\bibinfo{date}{Jan.} \bibinfo{year}{2023}).
\newblock
\showISSN{2055-2076}
\urldef\tempurl%
\url{https://doi.org/10.1177/20552076231180523}
\showDOI{\tempurl}


\bibitem[Guntuku et~al\mbox{.}(2019)]%
        {guntuku2019twitter}
\bibfield{author}{\bibinfo{person}{Sharath~Chandra Guntuku}, \bibinfo{person}{Daniel Preotiuc-Pietro}, \bibinfo{person}{Johannes~C Eichstaedt}, {and} \bibinfo{person}{Lyle~H Ungar}.} \bibinfo{year}{2019}\natexlab{}.
\newblock \showarticletitle{What twitter profile and posted images reveal about depression and anxiety}. In \bibinfo{booktitle}{\emph{Proceedings of the international AAAI conference on web and social media}}, Vol.~\bibinfo{volume}{13}. \bibinfo{pages}{236--246}.
\newblock


\bibitem[Guo et~al\mbox{.}(2021)]%
        {Guo2021}
\bibfield{author}{\bibinfo{person}{Weitong Guo}, \bibinfo{person}{Hongwu Yang}, \bibinfo{person}{Zhenyu Liu}, \bibinfo{person}{Yaping Xu}, {and} \bibinfo{person}{Bin Hu}.} \bibinfo{year}{2021}\natexlab{}.
\newblock \showarticletitle{Deep Neural Networks for Depression Recognition Based on 2D and 3D Facial Expressions Under Emotional Stimulus Tasks}.
\newblock \bibinfo{journal}{\emph{Frontiers in Neuroscience}}  \bibinfo{volume}{15} (\bibinfo{date}{April} \bibinfo{year}{2021}).
\newblock
\urldef\tempurl%
\url{https://doi.org/10.3389/fnins.2021.609760}
\showDOI{\tempurl}


\bibitem[Hosmer~Jr et~al\mbox{.}(2013)]%
        {hosmer2013applied}
\bibfield{author}{\bibinfo{person}{David~W Hosmer~Jr}, \bibinfo{person}{Stanley Lemeshow}, {and} \bibinfo{person}{Rodney~X Sturdivant}.} \bibinfo{year}{2013}\natexlab{}.
\newblock \bibinfo{booktitle}{\emph{Applied logistic regression}}. Vol.~\bibinfo{volume}{398}.
\newblock \bibinfo{publisher}{John Wiley \& Sons}.
\newblock


\bibitem[Hunt et~al\mbox{.}(2003)]%
        {Hunt2003}
\bibfield{author}{\bibinfo{person}{Melissa Hunt}, \bibinfo{person}{Joseph Auriemma}, {and} \bibinfo{person}{Ashara C.~A. Cashaw}.} \bibinfo{year}{2003}\natexlab{}.
\newblock \showarticletitle{Self-Report Bias and Underreporting of Depression on the {BDI}-{II}}.
\newblock \bibinfo{journal}{\emph{Journal of Personality Assessment}} \bibinfo{volume}{80}, \bibinfo{number}{1} (\bibinfo{date}{Feb.} \bibinfo{year}{2003}), \bibinfo{pages}{26--30}.
\newblock
\urldef\tempurl%
\url{https://doi.org/10.1207/s15327752jpa8001_10}
\showDOI{\tempurl}


\bibitem[Ikematsu et~al\mbox{.}(2020)]%
        {ikematsu2020investigating}
\bibfield{author}{\bibinfo{person}{Kaori Ikematsu}, \bibinfo{person}{Haruna Oshima}, \bibinfo{person}{Rachel Eardley}, {and} \bibinfo{person}{Itiro Siio}.} \bibinfo{year}{2020}\natexlab{}.
\newblock \showarticletitle{Investigating How Smartphone Movement is Affected by Lying Down Body Posture}.
\newblock \bibinfo{journal}{\emph{Proceedings of the ACM on Human-Computer Interaction}} \bibinfo{volume}{4}, \bibinfo{number}{ISS} (\bibinfo{year}{2020}), \bibinfo{pages}{1--17}.
\newblock


\bibitem[Joshi and Kanoongo(2022)]%
        {joshi2022depression}
\bibfield{author}{\bibinfo{person}{Manju~Lata Joshi} {and} \bibinfo{person}{Nehal Kanoongo}.} \bibinfo{year}{2022}\natexlab{}.
\newblock \showarticletitle{Depression detection using emotional artificial intelligence and machine learning: A closer review}.
\newblock \bibinfo{journal}{\emph{Materials Today: Proceedings}}  \bibinfo{volume}{58} (\bibinfo{year}{2022}), \bibinfo{pages}{217--226}.
\newblock


\bibitem[Khamis et~al\mbox{.}(2018)]%
        {Khamis2018}
\bibfield{author}{\bibinfo{person}{Mohamed Khamis}, \bibinfo{person}{Anita Baier}, \bibinfo{person}{Niels Henze}, \bibinfo{person}{Florian Alt}, {and} \bibinfo{person}{Andreas Bulling}.} \bibinfo{year}{2018}\natexlab{}.
\newblock \showarticletitle{Understanding Face and Eye Visibility in Front-Facing Cameras of Smartphones used in the Wild}. In \bibinfo{booktitle}{\emph{Proceedings of the 2018 {CHI} Conference on Human Factors in Computing Systems}}. \bibinfo{publisher}{{ACM}}.
\newblock
\urldef\tempurl%
\url{https://doi.org/10.1145/3173574.3173854}
\showDOI{\tempurl}


\bibitem[Kobak(2010)]%
        {Kobak2010}
\bibfield{author}{\bibinfo{person}{Kenneth~A. Kobak}.} \bibinfo{year}{2010}\natexlab{}.
\newblock \bibinfo{title}{Hamilton Depression Rating Scale}.
\newblock , \bibinfo{numpages}{1}~pages.
\newblock
\urldef\tempurl%
\url{https://doi.org/10.1002/9780470479216.corpsy0402}
\showDOI{\tempurl}


\bibitem[Kong et~al\mbox{.}(2022)]%
        {Kong2022}
\bibfield{author}{\bibinfo{person}{Xinru Kong}, \bibinfo{person}{Yan Yao}, \bibinfo{person}{Cuiying Wang}, \bibinfo{person}{Yuangeng Wang}, \bibinfo{person}{Jing Teng}, {and} \bibinfo{person}{Xianghua Qi}.} \bibinfo{year}{2022}\natexlab{}.
\newblock \showarticletitle{Automatic Identification of Depression Using Facial Images with Deep Convolutional Neural Network}.
\newblock \bibinfo{journal}{\emph{Medical Science Monitor}}  \bibinfo{volume}{28} (\bibinfo{date}{June} \bibinfo{year}{2022}).
\newblock
\urldef\tempurl%
\url{https://doi.org/10.12659/msm.936409}
\showDOI{\tempurl}


\bibitem[Kroenke et~al\mbox{.}(2001)]%
        {kroenke2001phq}
\bibfield{author}{\bibinfo{person}{Kurt Kroenke}, \bibinfo{person}{Robert~L Spitzer}, {and} \bibinfo{person}{Janet~BW Williams}.} \bibinfo{year}{2001}\natexlab{}.
\newblock \showarticletitle{The PHQ-9: validity of a brief depression severity measure}.
\newblock \bibinfo{journal}{\emph{Journal of general internal medicine}} \bibinfo{volume}{16}, \bibinfo{number}{9} (\bibinfo{year}{2001}), \bibinfo{pages}{606--613}.
\newblock


\bibitem[Kroenke et~al\mbox{.}(2010)]%
        {Kroenke2010-qi}
\bibfield{author}{\bibinfo{person}{Kurt Kroenke}, \bibinfo{person}{Robert~L Spitzer}, \bibinfo{person}{Janet B~W Williams}, {and} \bibinfo{person}{Bernd L{\"o}we}.} \bibinfo{year}{2010}\natexlab{}.
\newblock \showarticletitle{The Patient Health Questionnaire Somatic, Anxiety, and Depressive Symptom Scales: a systematic review}.
\newblock \bibinfo{journal}{\emph{Gen. Hosp. Psychiatry}} \bibinfo{volume}{32}, \bibinfo{number}{4} (\bibinfo{date}{July} \bibinfo{year}{2010}), \bibinfo{pages}{345--359}.
\newblock


\bibitem[Kroenke et~al\mbox{.}(2009)]%
        {kroenke2009phq}
\bibfield{author}{\bibinfo{person}{Kurt Kroenke}, \bibinfo{person}{Tara~W Strine}, \bibinfo{person}{Robert~L Spitzer}, \bibinfo{person}{Janet~BW Williams}, \bibinfo{person}{Joyce~T Berry}, {and} \bibinfo{person}{Ali~H Mokdad}.} \bibinfo{year}{2009}\natexlab{}.
\newblock \showarticletitle{The PHQ-8 as a measure of current depression in the general population}.
\newblock \bibinfo{journal}{\emph{Journal of affective disorders}} \bibinfo{volume}{114}, \bibinfo{number}{1-3} (\bibinfo{year}{2009}), \bibinfo{pages}{163--173}.
\newblock


\bibitem[Kurniawan et~al\mbox{.}(2020)]%
        {kurniawan2020design}
\bibfield{author}{\bibinfo{person}{Andreas Kurniawan}, \bibinfo{person}{Nunnun Bonafix}, \bibinfo{person}{Hendri Hartono}, {et~al\mbox{.}}} \bibinfo{year}{2020}\natexlab{}.
\newblock \showarticletitle{Design UI/UX Mobile Games for Left Hand Dominant People}.
\newblock \bibinfo{journal}{\emph{Journal of Games, Game Art, and Gamification}} \bibinfo{volume}{5}, \bibinfo{number}{2} (\bibinfo{year}{2020}), \bibinfo{pages}{48--53}.
\newblock


\bibitem[Lee and Park(2022)]%
        {Lee2022}
\bibfield{author}{\bibinfo{person}{Young-Shin Lee} {and} \bibinfo{person}{Won-Hyung Park}.} \bibinfo{year}{2022}\natexlab{}.
\newblock \showarticletitle{Diagnosis of Depressive Disorder Model on Facial Expression Based on Fast R-{CNN}}.
\newblock \bibinfo{journal}{\emph{Diagnostics}} \bibinfo{volume}{12}, \bibinfo{number}{2} (\bibinfo{date}{Jan.} \bibinfo{year}{2022}), \bibinfo{pages}{317}.
\newblock
\urldef\tempurl%
\url{https://doi.org/10.3390/diagnostics12020317}
\showDOI{\tempurl}


\bibitem[Levis et~al\mbox{.}(2019)]%
        {phq2019}
\bibfield{author}{\bibinfo{person}{Brooke Levis}, \bibinfo{person}{Andrea Benedetti}, {and} \bibinfo{person}{Brett~D Thombs}.} \bibinfo{year}{2019}\natexlab{}.
\newblock \showarticletitle{Accuracy of Patient Health Questionnaire-9 (PHQ-9) for screening to detect major depression: individual participant data meta-analysis}.
\newblock \bibinfo{journal}{\emph{BMJ}} (\bibinfo{date}{April} \bibinfo{year}{2019}), \bibinfo{pages}{l1476}.
\newblock
\showISSN{1756-1833}
\urldef\tempurl%
\url{https://doi.org/10.1136/bmj.l1476}
\showDOI{\tempurl}


\bibitem[Li et~al\mbox{.}(2022)]%
        {li2022blip}
\bibfield{author}{\bibinfo{person}{Junnan Li}, \bibinfo{person}{Dongxu Li}, \bibinfo{person}{Caiming Xiong}, {and} \bibinfo{person}{Steven Hoi}.} \bibinfo{year}{2022}\natexlab{}.
\newblock \showarticletitle{Blip: Bootstrapping language-image pre-training for unified vision-language understanding and generation}. In \bibinfo{booktitle}{\emph{International Conference on Machine Learning}}. PMLR, \bibinfo{pages}{12888--12900}.
\newblock


\bibitem[Liu et~al\mbox{.}(2022)]%
        {Liu2022}
\bibfield{author}{\bibinfo{person}{Dongdong Liu}, \bibinfo{person}{Bowen Liu}, \bibinfo{person}{Tao Lin}, \bibinfo{person}{Guangya Liu}, \bibinfo{person}{Guoyu Yang}, \bibinfo{person}{Dezhen Qi}, \bibinfo{person}{Ye Qiu}, \bibinfo{person}{Yuer Lu}, \bibinfo{person}{Qinmei Yuan}, \bibinfo{person}{Stella~C. Shuai}, \bibinfo{person}{Xiang Li}, \bibinfo{person}{Ou Liu}, \bibinfo{person}{Xiangdong Tang}, \bibinfo{person}{Jianwei Shuai}, \bibinfo{person}{Yuping Cao}, {and} \bibinfo{person}{Hai Lin}.} \bibinfo{year}{2022}\natexlab{}.
\newblock \showarticletitle{Measuring depression severity based on facial expression and body movement using deep convolutional neural network}.
\newblock \bibinfo{journal}{\emph{Frontiers in Psychiatry}}  \bibinfo{volume}{13} (\bibinfo{date}{Dec.} \bibinfo{year}{2022}).
\newblock
\urldef\tempurl%
\url{https://doi.org/10.3389/fpsyt.2022.1017064}
\showDOI{\tempurl}


\bibitem[L{\"o}bner et~al\mbox{.}(2023)]%
        {lobner2023user}
\bibfield{author}{\bibinfo{person}{Sascha L{\"o}bner}, \bibinfo{person}{Sebastian Pape}, {and} \bibinfo{person}{Vanessa Bracamonte}.} \bibinfo{year}{2023}\natexlab{}.
\newblock \showarticletitle{User Acceptance Criteria for Privacy Preserving Machine Learning Techniques}. In \bibinfo{booktitle}{\emph{Proceedings of the 18th International Conference on Availability, Reliability and Security}}. \bibinfo{pages}{1--8}.
\newblock


\bibitem[Lum et~al\mbox{.}(2016)]%
        {Lum2016}
\bibfield{author}{\bibinfo{person}{Hillary~D. Lum}, \bibinfo{person}{Evan~P. Carey}, \bibinfo{person}{Diane Fairclough}, \bibinfo{person}{Mary~E. Plomondon}, \bibinfo{person}{Evelyn Hutt}, \bibinfo{person}{John~S. Rumsfeld}, {and} \bibinfo{person}{David~B. Bekelman}.} \bibinfo{year}{2016}\natexlab{}.
\newblock \showarticletitle{Burdensome Physical and Depressive Symptoms Predict Heart Failure–Specific Health Status Over One Year}.
\newblock \bibinfo{journal}{\emph{Journal of Pain and Symptom Management}} \bibinfo{volume}{51}, \bibinfo{number}{6} (\bibinfo{date}{June} \bibinfo{year}{2016}), \bibinfo{pages}{963–970}.
\newblock
\showISSN{0885-3924}
\urldef\tempurl%
\url{https://doi.org/10.1016/j.jpainsymman.2015.12.328}
\showDOI{\tempurl}


\bibitem[Lundberg and Lee(2017)]%
        {lundberg2017unified}
\bibfield{author}{\bibinfo{person}{Scott~M Lundberg} {and} \bibinfo{person}{Su-In Lee}.} \bibinfo{year}{2017}\natexlab{}.
\newblock \showarticletitle{A unified approach to interpreting model predictions}.
\newblock \bibinfo{journal}{\emph{Advances in neural information processing systems}}  \bibinfo{volume}{30} (\bibinfo{year}{2017}).
\newblock


\bibitem[Majethia et~al\mbox{.}(2022)]%
        {Majethia2022}
\bibfield{author}{\bibinfo{person}{Rahul Majethia}, \bibinfo{person}{Vadlamudi~Pratiksha Sharma}, {and} \bibinfo{person}{Rishika Dwaraghanath}.} \bibinfo{year}{2022}\natexlab{}.
\newblock \showarticletitle{Mental Health Indices as Biomarkers for Assistive Mental Healthcare in University Students}. In \bibinfo{booktitle}{\emph{2022 10th International Conference on Affective Computing and Intelligent Interaction (ACII)}}. \bibinfo{publisher}{IEEE}.
\newblock
\urldef\tempurl%
\url{https://doi.org/10.1109/acii55700.2022.9953847}
\showDOI{\tempurl}


\bibitem[Mellouk and Handouzi(2020)]%
        {Mellouk2020}
\bibfield{author}{\bibinfo{person}{Wafa Mellouk} {and} \bibinfo{person}{Wahida Handouzi}.} \bibinfo{year}{2020}\natexlab{}.
\newblock \showarticletitle{Facial emotion recognition using deep learning: review and insights}.
\newblock \bibinfo{journal}{\emph{Procedia Computer Science}}  \bibinfo{volume}{175} (\bibinfo{year}{2020}), \bibinfo{pages}{689--694}.
\newblock
\urldef\tempurl%
\url{https://doi.org/10.1016/j.procs.2020.07.101}
\showDOI{\tempurl}


\bibitem[Nelavelli and Ploetz(2018)]%
        {nelavelli2018adaptive}
\bibfield{author}{\bibinfo{person}{Kriti Nelavelli} {and} \bibinfo{person}{Thomas Ploetz}.} \bibinfo{year}{2018}\natexlab{}.
\newblock \showarticletitle{Adaptive App Design by Detecting Handedness}.
\newblock \bibinfo{journal}{\emph{arXiv preprint arXiv:1805.08367}} (\bibinfo{year}{2018}).
\newblock


\bibitem[Nepal et~al\mbox{.}(2022)]%
        {10.1145/3491102.3502043}
\bibfield{author}{\bibinfo{person}{Subigya Nepal}, \bibinfo{person}{Weichen Wang}, \bibinfo{person}{Vlado Vojdanovski}, \bibinfo{person}{Jeremy~F Huckins}, \bibinfo{person}{Alex daSilva}, \bibinfo{person}{Meghan Meyer}, {and} \bibinfo{person}{Andrew Campbell}.} \bibinfo{year}{2022}\natexlab{}.
\newblock \showarticletitle{COVID Student Study: A Year in the Life of College Students during the COVID-19 Pandemic Through the Lens of Mobile Phone Sensing}. In \bibinfo{booktitle}{\emph{Proceedings of the 2022 CHI Conference on Human Factors in Computing Systems}} (New Orleans, LA, USA) \emph{(\bibinfo{series}{CHI '22})}. \bibinfo{publisher}{Association for Computing Machinery}, \bibinfo{address}{New York, NY, USA}, Article \bibinfo{articleno}{42}, \bibinfo{numpages}{19}~pages.
\newblock
\showISBNx{9781450391573}
\urldef\tempurl%
\url{https://doi.org/10.1145/3491102.3502043}
\showDOI{\tempurl}


\bibitem[Nguyen et~al\mbox{.}(2021)]%
        {Nguyen2021}
\bibfield{author}{\bibinfo{person}{Minh~X Nguyen}, \bibinfo{person}{H.~Luz McNaughton~Reyes}, \bibinfo{person}{Brian~W Pence}, \bibinfo{person}{Kate Muessig}, \bibinfo{person}{Heidi~E Hutton}, \bibinfo{person}{Carl~A Latkin}, \bibinfo{person}{David Dowdy}, \bibinfo{person}{Geetanjali Chander}, \bibinfo{person}{Kathryn~E Lancaster}, \bibinfo{person}{Constantine Frangakis}, \bibinfo{person}{Teerada Sripaipan}, \bibinfo{person}{Viet Ha~Tran}, {and} \bibinfo{person}{Vivian~F Go}.} \bibinfo{year}{2021}\natexlab{}.
\newblock \showarticletitle{The longitudinal association between depression, anxiety symptoms and HIV outcomes, and the modifying effect of alcohol dependence among ART clients with hazardous alcohol use in Vietnam}.
\newblock \bibinfo{journal}{\emph{Journal of the International AIDS Society}} \bibinfo{volume}{24}, \bibinfo{number}{S2} (\bibinfo{date}{June} \bibinfo{year}{2021}).
\newblock
\showISSN{1758-2652}
\urldef\tempurl%
\url{https://doi.org/10.1002/jia2.25746}
\showDOI{\tempurl}


\bibitem[Olver and Hopwood(2013)]%
        {Olver2013}
\bibfield{author}{\bibinfo{person}{James~S Olver} {and} \bibinfo{person}{Malcolm~J Hopwood}.} \bibinfo{year}{2013}\natexlab{}.
\newblock \showarticletitle{Depression and physical illness}.
\newblock \bibinfo{journal}{\emph{Medical Journal of Australia}} \bibinfo{volume}{199}, \bibinfo{number}{S6} (\bibinfo{date}{Oct.} \bibinfo{year}{2013}).
\newblock
\urldef\tempurl%
\url{https://doi.org/10.5694/mja12.10597}
\showDOI{\tempurl}


\bibitem[Organization(2023)]%
        {WHO2023}
\bibfield{author}{\bibinfo{person}{World~Health Organization}.} \bibinfo{year}{2023}\natexlab{}.
\newblock \bibinfo{title}{Mental disorders}.
\newblock
\newblock
\urldef\tempurl%
\url{https://www.who.int/news-room/fact-sheets/detail/mental-disorders}
\showURL{%
\tempurl}
\newblock
\shownote{Accessed: [2023]}.


\bibitem[Pampouchidou et~al\mbox{.}(2015)]%
        {7412257}
\bibfield{author}{\bibinfo{person}{A. Pampouchidou}, \bibinfo{person}{K. Marias}, \bibinfo{person}{M. Tsiknakis}, \bibinfo{person}{P. Simos}, \bibinfo{person}{F. Yang}, {and} \bibinfo{person}{F. Meriaudeau}.} \bibinfo{year}{2015}\natexlab{}.
\newblock \showarticletitle{Designing a framework for assisting depression severity assessment from facial image analysis}. In \bibinfo{booktitle}{\emph{2015 IEEE International Conference on Signal and Image Processing Applications (ICSIPA)}}. \bibinfo{pages}{578--583}.
\newblock
\urldef\tempurl%
\url{https://doi.org/10.1109/ICSIPA.2015.7412257}
\showDOI{\tempurl}


\bibitem[Pampouchidou et~al\mbox{.}(2017)]%
        {pampouchidou2017facial}
\bibfield{author}{\bibinfo{person}{Anastasia Pampouchidou}, \bibinfo{person}{Olympia Simantiraki}, \bibinfo{person}{C-M Vazakopoulou}, \bibinfo{person}{Charikleia Chatzaki}, \bibinfo{person}{Matthew Pediaditis}, \bibinfo{person}{Anna Maridaki}, \bibinfo{person}{Kostas Marias}, \bibinfo{person}{Panagiotis Simos}, \bibinfo{person}{Fan Yang}, \bibinfo{person}{Fabrice Meriaudeau}, {et~al\mbox{.}}} \bibinfo{year}{2017}\natexlab{}.
\newblock \showarticletitle{Facial geometry and speech analysis for depression detection}. In \bibinfo{booktitle}{\emph{2017 39th Annual International Conference of the IEEE Engineering in Medicine and Biology Society (EMBC)}}. IEEE, \bibinfo{pages}{1433--1436}.
\newblock


\bibitem[Pillai et~al\mbox{.}(2024)]%
        {pillai2024investigating}
\bibfield{author}{\bibinfo{person}{Arvind Pillai}, \bibinfo{person}{Subigya~Kumar Nepal}, \bibinfo{person}{Weichen Wang}, \bibinfo{person}{Matthew Nemesure}, \bibinfo{person}{Michael Heinz}, \bibinfo{person}{George Price}, \bibinfo{person}{Damien Lekkas}, \bibinfo{person}{Amanda~C Collins}, \bibinfo{person}{Tess Griffin}, \bibinfo{person}{Benjamin Buck}, {et~al\mbox{.}}} \bibinfo{year}{2024}\natexlab{}.
\newblock \showarticletitle{Investigating Generalizability of Speech-based Suicidal Ideation Detection Using Mobile Phones}.
\newblock \bibinfo{journal}{\emph{Proceedings of the ACM on Interactive, Mobile, Wearable and Ubiquitous Technologies}} \bibinfo{volume}{7}, \bibinfo{number}{4} (\bibinfo{year}{2024}), \bibinfo{pages}{1--38}.
\newblock


\bibitem[Ramos-Cuadros et~al\mbox{.}(2021)]%
        {ramos2021evaluating}
\bibfield{author}{\bibinfo{person}{Alexander Ramos-Cuadros}, \bibinfo{person}{Luis~Palomino Santillan}, {and} \bibinfo{person}{Willy Ugarte}.} \bibinfo{year}{2021}\natexlab{}.
\newblock \showarticletitle{Evaluating the Depression Level Based on Facial Image Analyzing and Patient Voice}. In \bibinfo{booktitle}{\emph{International Conference on Information and Communication Technologies for Ageing Well and e-Health}}. Springer, \bibinfo{pages}{35--55}.
\newblock


\bibitem[Reece and Danforth(2017)]%
        {Reece2017}
\bibfield{author}{\bibinfo{person}{Andrew~G Reece} {and} \bibinfo{person}{Christopher~M Danforth}.} \bibinfo{year}{2017}\natexlab{}.
\newblock \showarticletitle{Instagram photos reveal predictive markers of depression}.
\newblock \bibinfo{journal}{\emph{{EPJ} Data Science}} \bibinfo{volume}{6}, \bibinfo{number}{1} (\bibinfo{date}{Aug.} \bibinfo{year}{2017}).
\newblock
\urldef\tempurl%
\url{https://doi.org/10.1140/epjds/s13688-017-0110-z}
\showDOI{\tempurl}


\bibitem[Rottenberg et~al\mbox{.}(2005)]%
        {Rottenberg2005}
\bibfield{author}{\bibinfo{person}{Jonathan Rottenberg}, \bibinfo{person}{James~J. Gross}, {and} \bibinfo{person}{Ian~H. Gotlib}.} \bibinfo{year}{2005}\natexlab{}.
\newblock \showarticletitle{Emotion Context Insensitivity in Major Depressive Disorder.}
\newblock \bibinfo{journal}{\emph{Journal of Abnormal Psychology}} \bibinfo{volume}{114}, \bibinfo{number}{4} (\bibinfo{date}{Nov.} \bibinfo{year}{2005}), \bibinfo{pages}{627--639}.
\newblock
\urldef\tempurl%
\url{https://doi.org/10.1037/0021-843x.114.4.627}
\showDOI{\tempurl}


\bibitem[Saarni et~al\mbox{.}(2007)]%
        {Saarni2007-sd}
\bibfield{author}{\bibinfo{person}{Samuli~I Saarni}, \bibinfo{person}{Jaana Suvisaari}, \bibinfo{person}{Harri Sintonen}, \bibinfo{person}{Sami Pirkola}, \bibinfo{person}{Seppo Koskinen}, \bibinfo{person}{Arpo Aromaa}, {and} \bibinfo{person}{Jouko L{\"o}nnqvist}.} \bibinfo{year}{2007}\natexlab{}.
\newblock \showarticletitle{Impact of psychiatric disorders on health-related quality of life: general population survey}.
\newblock \bibinfo{journal}{\emph{Br. J. Psychiatry}}  \bibinfo{volume}{190} (\bibinfo{date}{April} \bibinfo{year}{2007}), \bibinfo{pages}{326--332}.
\newblock


\bibitem[Saggu et~al\mbox{.}(2022)]%
        {saggu2022depressnet}
\bibfield{author}{\bibinfo{person}{Guramritpal~Singh Saggu}, \bibinfo{person}{Keshav Gupta}, \bibinfo{person}{KV Arya}, {and} \bibinfo{person}{Ciro~Rodriguez Rodriguez}.} \bibinfo{year}{2022}\natexlab{}.
\newblock \showarticletitle{DepressNet: A Multimodal Hierarchical Attention Mechanism approach for Depression Detection}.
\newblock \bibinfo{journal}{\emph{Int. J. Eng. Sci.}} \bibinfo{volume}{15}, \bibinfo{number}{1} (\bibinfo{year}{2022}), \bibinfo{pages}{24--32}.
\newblock


\bibitem[Sagonas et~al\mbox{.}(2016)]%
        {sagonas2016300}
\bibfield{author}{\bibinfo{person}{Christos Sagonas}, \bibinfo{person}{Epameinondas Antonakos}, \bibinfo{person}{Georgios Tzimiropoulos}, \bibinfo{person}{Stefanos Zafeiriou}, {and} \bibinfo{person}{Maja Pantic}.} \bibinfo{year}{2016}\natexlab{}.
\newblock \showarticletitle{300 faces in-the-wild challenge: Database and results}.
\newblock \bibinfo{journal}{\emph{Image and vision computing}}  \bibinfo{volume}{47} (\bibinfo{year}{2016}), \bibinfo{pages}{3--18}.
\newblock


\bibitem[Sagonas et~al\mbox{.}(2013)]%
        {sagonas2013300}
\bibfield{author}{\bibinfo{person}{Christos Sagonas}, \bibinfo{person}{Georgios Tzimiropoulos}, \bibinfo{person}{Stefanos Zafeiriou}, {and} \bibinfo{person}{Maja Pantic}.} \bibinfo{year}{2013}\natexlab{}.
\newblock \showarticletitle{300 faces in-the-wild challenge: The first facial landmark localization challenge}. In \bibinfo{booktitle}{\emph{Proceedings of the IEEE international conference on computer vision workshops}}. \bibinfo{pages}{397--403}.
\newblock


\bibitem[Santini et~al\mbox{.}(2015)]%
        {Santini2015}
\bibfield{author}{\bibinfo{person}{Ziggi~Ivan Santini}, \bibinfo{person}{Ai Koyanagi}, \bibinfo{person}{Stefanos Tyrovolas}, \bibinfo{person}{Catherine Mason}, {and} \bibinfo{person}{Josep~Maria Haro}.} \bibinfo{year}{2015}\natexlab{}.
\newblock \showarticletitle{The association between social relationships and depression: A systematic review}.
\newblock \bibinfo{journal}{\emph{Journal of Affective Disorders}}  \bibinfo{volume}{175} (\bibinfo{date}{April} \bibinfo{year}{2015}), \bibinfo{pages}{53--65}.
\newblock
\urldef\tempurl%
\url{https://doi.org/10.1016/j.jad.2014.12.049}
\showDOI{\tempurl}


\bibitem[Sarsenbayeva et~al\mbox{.}(2017)]%
        {sarsenbayeva2017challenges}
\bibfield{author}{\bibinfo{person}{Zhanna Sarsenbayeva}, \bibinfo{person}{Niels van Berkel}, \bibinfo{person}{Chu Luo}, \bibinfo{person}{Vassilis Kostakos}, {and} \bibinfo{person}{Jorge Goncalves}.} \bibinfo{year}{2017}\natexlab{}.
\newblock \showarticletitle{Challenges of situational impairments during interaction with mobile devices}. In \bibinfo{booktitle}{\emph{Proceedings of the 29th australian conference on computer-human interaction}}. \bibinfo{pages}{477--481}.
\newblock


\bibitem[Schomerus et~al\mbox{.}(2012)]%
        {Schomerus2012}
\bibfield{author}{\bibinfo{person}{Georg Schomerus}, \bibinfo{person}{Charlotte Auer}, \bibinfo{person}{Dieter Rhode}, \bibinfo{person}{Melanie Luppa}, \bibinfo{person}{Harald~J. Freyberger}, {and} \bibinfo{person}{Silke Schmidt}.} \bibinfo{year}{2012}\natexlab{}.
\newblock \showarticletitle{Personal stigma, problem appraisal and perceived need for professional help in currently untreated depressed persons}.
\newblock \bibinfo{journal}{\emph{Journal of Affective Disorders}} \bibinfo{volume}{139}, \bibinfo{number}{1} (\bibinfo{date}{June} \bibinfo{year}{2012}), \bibinfo{pages}{94--97}.
\newblock
\urldef\tempurl%
\url{https://doi.org/10.1016/j.jad.2012.02.022}
\showDOI{\tempurl}


\bibitem[Tan and Le(2019)]%
        {tan2019efficientnet}
\bibfield{author}{\bibinfo{person}{Mingxing Tan} {and} \bibinfo{person}{Quoc Le}.} \bibinfo{year}{2019}\natexlab{}.
\newblock \showarticletitle{Efficientnet: Rethinking model scaling for convolutional neural networks}. In \bibinfo{booktitle}{\emph{International conference on machine learning}}. PMLR, \bibinfo{pages}{6105--6114}.
\newblock


\bibitem[Tigwell et~al\mbox{.}(2018)]%
        {tigwell2018s}
\bibfield{author}{\bibinfo{person}{Garreth~W Tigwell}, \bibinfo{person}{David~R Flatla}, {and} \bibinfo{person}{Rachel Menzies}.} \bibinfo{year}{2018}\natexlab{}.
\newblock \showarticletitle{It's not just the light: understanding the factors causing situational visual impairments during mobile interaction}. In \bibinfo{booktitle}{\emph{Proceedings of the 10th Nordic Conference on Human-Computer Interaction}}. \bibinfo{pages}{338--351}.
\newblock


\bibitem[Valdez and Mehrabian(1994)]%
        {valdez1994effects}
\bibfield{author}{\bibinfo{person}{Patricia Valdez} {and} \bibinfo{person}{Albert Mehrabian}.} \bibinfo{year}{1994}\natexlab{}.
\newblock \showarticletitle{Effects of color on emotions.}
\newblock \bibinfo{journal}{\emph{Journal of experimental psychology: General}} \bibinfo{volume}{123}, \bibinfo{number}{4} (\bibinfo{year}{1994}), \bibinfo{pages}{394}.
\newblock


\bibitem[Wang et~al\mbox{.}(2015)]%
        {10.1145/2800835.2804391}
\bibfield{author}{\bibinfo{person}{Rui Wang}, \bibinfo{person}{Andrew~T. Campbell}, {and} \bibinfo{person}{Xia Zhou}.} \bibinfo{year}{2015}\natexlab{}.
\newblock \showarticletitle{Using Opportunistic Face Logging from Smartphone to Infer Mental Health: Challenges and Future Directions}. In \bibinfo{booktitle}{\emph{Adjunct Proceedings of the 2015 ACM International Joint Conference on Pervasive and Ubiquitous Computing and Proceedings of the 2015 ACM International Symposium on Wearable Computers}} (Osaka, Japan) \emph{(\bibinfo{series}{UbiComp/ISWC'15 Adjunct})}. \bibinfo{publisher}{Association for Computing Machinery}, \bibinfo{address}{New York, NY, USA}, \bibinfo{pages}{683–692}.
\newblock
\showISBNx{9781450335751}
\urldef\tempurl%
\url{https://doi.org/10.1145/2800835.2804391}
\showDOI{\tempurl}


\bibitem[Wang et~al\mbox{.}(2014)]%
        {wang2014studentlife}
\bibfield{author}{\bibinfo{person}{Rui Wang}, \bibinfo{person}{Fanglin Chen}, \bibinfo{person}{Zhenyu Chen}, \bibinfo{person}{Tianxing Li}, \bibinfo{person}{Gabriella Harari}, \bibinfo{person}{Stefanie Tignor}, \bibinfo{person}{Xia Zhou}, \bibinfo{person}{Dror Ben-Zeev}, {and} \bibinfo{person}{Andrew~T Campbell}.} \bibinfo{year}{2014}\natexlab{}.
\newblock \showarticletitle{StudentLife: assessing mental health, academic performance and behavioral trends of college students using smartphones}. In \bibinfo{booktitle}{\emph{Proceedings of the 2014 ACM international joint conference on pervasive and ubiquitous computing}}. \bibinfo{pages}{3--14}.
\newblock


\bibitem[Wang et~al\mbox{.}(2018)]%
        {wang2018tracking}
\bibfield{author}{\bibinfo{person}{Rui Wang}, \bibinfo{person}{Weichen Wang}, \bibinfo{person}{Alex DaSilva}, \bibinfo{person}{Jeremy~F Huckins}, \bibinfo{person}{William~M Kelley}, \bibinfo{person}{Todd~F Heatherton}, {and} \bibinfo{person}{Andrew~T Campbell}.} \bibinfo{year}{2018}\natexlab{}.
\newblock \showarticletitle{Tracking depression dynamics in college students using mobile phone and wearable sensing}.
\newblock \bibinfo{journal}{\emph{Proceedings of the ACM on Interactive, Mobile, Wearable and Ubiquitous Technologies}} \bibinfo{volume}{2}, \bibinfo{number}{1} (\bibinfo{year}{2018}), \bibinfo{pages}{1--26}.
\newblock


\bibitem[Wood et~al\mbox{.}(2015)]%
        {wood2015rendering}
\bibfield{author}{\bibinfo{person}{Erroll Wood}, \bibinfo{person}{Tadas Baltrusaitis}, \bibinfo{person}{Xucong Zhang}, \bibinfo{person}{Yusuke Sugano}, \bibinfo{person}{Peter Robinson}, {and} \bibinfo{person}{Andreas Bulling}.} \bibinfo{year}{2015}\natexlab{}.
\newblock \showarticletitle{Rendering of eyes for eye-shape registration and gaze estimation}. In \bibinfo{booktitle}{\emph{Proceedings of the IEEE international conference on computer vision}}. \bibinfo{pages}{3756--3764}.
\newblock


\bibitem[Xu et~al\mbox{.}(2019)]%
        {xu2019leveraging}
\bibfield{author}{\bibinfo{person}{Xuhai Xu}, \bibinfo{person}{Prerna Chikersal}, \bibinfo{person}{Afsaneh Doryab}, \bibinfo{person}{Daniella~K Villalba}, \bibinfo{person}{Janine~M Dutcher}, \bibinfo{person}{Michael~J Tumminia}, \bibinfo{person}{Tim Althoff}, \bibinfo{person}{Sheldon Cohen}, \bibinfo{person}{Kasey~G Creswell}, \bibinfo{person}{J~David Creswell}, {et~al\mbox{.}}} \bibinfo{year}{2019}\natexlab{}.
\newblock \showarticletitle{Leveraging routine behavior and contextually-filtered features for depression detection among college students}.
\newblock \bibinfo{journal}{\emph{Proceedings of the ACM on Interactive, Mobile, Wearable and Ubiquitous Technologies}} \bibinfo{volume}{3}, \bibinfo{number}{3} (\bibinfo{year}{2019}), \bibinfo{pages}{1--33}.
\newblock


\bibitem[Zhou et~al\mbox{.}(2018)]%
        {zhou2018visually}
\bibfield{author}{\bibinfo{person}{Xiuzhuang Zhou}, \bibinfo{person}{Kai Jin}, \bibinfo{person}{Yuanyuan Shang}, {and} \bibinfo{person}{Guodong Guo}.} \bibinfo{year}{2018}\natexlab{}.
\newblock \showarticletitle{Visually interpretable representation learning for depression recognition from facial images}.
\newblock \bibinfo{journal}{\emph{IEEE transactions on affective computing}} \bibinfo{volume}{11}, \bibinfo{number}{3} (\bibinfo{year}{2018}), \bibinfo{pages}{542--552}.
\newblock


\bibitem[Zou and Hastie(2005)]%
        {zou2005regularization}
\bibfield{author}{\bibinfo{person}{Hui Zou} {and} \bibinfo{person}{Trevor Hastie}.} \bibinfo{year}{2005}\natexlab{}.
\newblock \showarticletitle{Regularization and variable selection via the elastic net}.
\newblock \bibinfo{journal}{\emph{Journal of the Royal Statistical Society Series B: Statistical Methodology}} \bibinfo{volume}{67}, \bibinfo{number}{2} (\bibinfo{year}{2005}), \bibinfo{pages}{301--320}.
\newblock


\end{thebibliography}
\clearpage
\appendix
\section{Surveys}
\label{sec:appxsurvey}
\begin{table}[h]
\smaller
\centering
\caption{PHQ-8 Questionnaire \cite{kroenke2001phq}}
\label{tab:phq8}
\begin{tabularx}{0.45\textwidth}{cX} 
\toprule
\textbf{No.} & \textbf{Question} \\
\midrule
& In the past 4 hours... \\
1 & I have had little interest or pleasure in doing things \\
2 & I have felt down, depressed, or hopeless \\
3 & Last night I had trouble with sleep\\
4 & I have felt tired or have had little energy \\
5 & I have had a poor appetite or have been overeating \\
6 & I have felt bad about myself\\
7 & I have had trouble concentrating\\
8 & I have been moving or speaking slowly, or fidgeting more. \\
\bottomrule
\end{tabularx}
\end{table}

\begin{table}[h]
\smaller
\centering
\caption{User Comfort Questionnaire}
\label{tab:comfort}
\begin{tabularx}{0.45\textwidth}{cX} 
\toprule
\textbf{No.} & \textbf{Question} \\
\midrule
1 & As you know, participating in this study is confidential and we have multiple measures in place to protect the data you've shared. Given that, we would like to know how comfortable you were with sharing the front-facing photo bursts while taking surveys \\
2 & Please tell us why the photo bursts made you uncomfortable. \\
\bottomrule
\end{tabularx}
\end{table}

\section{Additional Metrics for Models}
\label{sec:r2}
\begin{table}[h]
\caption{\textcolor{black}{Performance: R-squared ($\pmb{R^2}$) values for PHQ-8 regression score prediction. `LR + EN' refers to logistic regression for depression classification and elastic net for regression.}}
\small
\begin{tabular}{@{}ll@{}}
\toprule
\textbf{Method} &\textbf{$\pmb{R^2}$} \\ \midrule
Baseline & 0.05  \\ \cdashline{0-1}
LR + EN (MI) &0.12 \\
Random Forest (MI) & 0.14  \\
Random Forest (3D Landmarks) &0.20 \\ 
EffNet &0.13 \\
\bottomrule
\end{tabular}
\end{table}

\begin{table}[h]
\caption{\textcolor{black}{Ablation Study: Investigating R-squared ($\pmb{R^2}$) values for PHQ-8 regression score prediction of OpenFace feature sets using a \textit{random forest}.}}
\small
\begin{tabular}{@{}ll@{}}
\toprule
\textbf{Method} &\textbf{$\pmb{R^2}$} \\ \midrule
Facial Action Units &0.13  \\ \cdashline{0-1}
Gaze &0.11 \\
Eye Landmarks & 0.17  \\
Head Pose &0.16 \\ 
Rigidity Parameters &0.13 \\
2D Landmarks &0.16 \\
3D Landmarks &0.20 \\
\bottomrule
\end{tabular}
\end{table}

\end{document}